\documentclass[times,twocolumn]{aastex631}
\usepackage{CJK}
\usepackage{amsmath, amssymb, bm}
\usepackage{mathrsfs}
\usepackage[nodayofweek]{datetime}

\newcommand{\uat}[2]{\href{http://astrothesaurus.org/uat/#1}{#2  (#1)}}
\newcommand{\hi}{\ensuremath{\text{H\,{\sc i}}}}
\newcommand{\nhi}{\ensuremath{N_{\text{H\,{\sc i}}}}}

\newcommand{\shi}{\ensuremath{S_{\text{H\,{\sc i}}}}}

\usepackage{xcolor}

\makeatletter
\patchcmd{\NAT@citex}
  {\@citea\NAT@hyper@{%
     \NAT@nmfmt{\NAT@nm}%
     \hyper@natlinkbreak{\NAT@aysep\NAT@spacechar}{\@citeb\@extra@b@citeb}%
     \NAT@date}}
  {\@citea\NAT@nmfmt{\NAT@nm}%
   \NAT@aysep\NAT@spacechar\NAT@hyper@{\NAT@date}}{}{}

\patchcmd{\NAT@citex}
  {\@citea\NAT@hyper@{%
     \NAT@nmfmt{\NAT@nm}%
     \hyper@natlinkbreak{\NAT@spacechar\NAT@@open\if*#1*\else#1\NAT@spacechar\fi}%
       {\@citeb\@extra@b@citeb}%
     \NAT@date}}
  {\@citea\NAT@nmfmt{\NAT@nm}%
   \NAT@spacechar\NAT@@open\if*#1*\else#1\NAT@spacechar\fi\NAT@hyper@{\NAT@date}}
  {}{}
\makeatother

\shortauthors{Su et al.}

\begin{document}
\begin{CJK*}{UTF8}{gbsn}
\title{A FAST Survey of \hi\ Absorption in Low-power Radio Sources}

\correspondingauthor{Qingzheng Yu}
\email{qingzheng.yu@unifi.it}

\author{Yang Su (苏杨)}
\affiliation{Department of Astronomy, Xiamen University, Xiamen, Fujian 361005, People's Republic of China}

\author[0000-0003-3230-3981]{Qingzheng Yu (余清正)}
\affiliation{Dipartimento di Fisica e Astronomia, Universit\`a degli Studi di Firenze, Via G. Sansone 1, 50019 Sesto Fiorentino, Firenze, Italy}
\affiliation{INAF—Osservatorio Astrofisico di Arcetri, Largo E. Fermi 5, I-50125 Firenze, Italy}

\author[0000-0002-2853-3808]{Taotao Fang (方陶陶)}
\affiliation{Department of Astronomy, Xiamen University, Xiamen, Fujian 361005, People's Republic of China}

\author[0000-0003-4874-0369]{Junfeng Wang (王俊峰)}
\affiliation{Department of Astronomy, Xiamen University, Xiamen, Fujian 361005, People's Republic of China}

\author[0000-0001-7349-4695]{Jianfeng Wu (武剑锋)}
\affiliation{Department of Astronomy, Xiamen University, Xiamen, Fujian 361005, People's Republic of China}

\author{Bo Zhang (张博)}
\affiliation{National Astronomical Observatories, Chinese Academy of Sciences, 20A Datun Road, Chaoyang District, Beijing 100101, People's Republic of China}
\affiliation{Guizhou Radio Astronomical Observatory, Guiyang 550025, People's Republic of China}
\affiliation{Guizhou Provincial Key Laboratory of Medium and Low Frequency Radio Astronomy Technology and Application，Guiyang 550025, People's Republic of China}


\begin{abstract}

We conducted a \hi\ 21\,cm absorption study of a sample of 147 nearby ($z < 0.1$) low-power radio sources with $10\,\mathrm{mJy} < S_{1.4\,\mathrm{GHz}} < 30\,\mathrm{mJy}$ and $\log(P_{1.4\,\mathrm{GHz}}/\mathrm{W\,Hz^{-1}}) = 20.5-23.7$, using the Five-hundred-meter Aperture Spherical radio Telescope. By investigating the origin and kinematics of \hi\ absorbing gas, we aim to study the interplay between the active galactic nucleus (AGN) and its surrounding interstellar medium. Our observations detect 12 new absorbers, combining results from the pilot survey (three absorbers out of 26 sources), yielding a detection rate of $\sim10.2^{+3.1}_{-2.0}\%$. The detection rate in our sample is lower than in higher-power samples, which is likely due to emission dilution and the dominance of extended sources, indicating a gas-rich and star-forming-dominated population in low-power sources. Among new detections, most line profiles are narrow and show velocities close to systemic ones, consistent with rotating disks, while four show disturbed kinematics indicative of inflows or outflows. The fraction of outflow candidates rises with radio power, while the fraction of inflow ones remains constant, suggesting the effect of radio emission on driving \hi\ outflows. In our sample, compact sources show a higher \hi\ detection rate than extended sources. Contrary to expectations from higher-power samples, MIR-bright sources at low-power radio do not exhibit a  higher \hi\ detection rate or more disturbed kinematics. In low-power radio sources, blueshifted absorption occurs only in Seyferts and low-ionization nuclear emitting
regions, indicating the connection between atomic outflows and the ionization state of AGN.

\end{abstract}

\keywords{\uat{16}{Active galactic nuclei}; \uat{2017}{AGN host galaxies}; \uat{847}{Interstellar medium}; \uat{1099}{Neutral hydrogen clouds}; \uat{508}{Extragalactic radio sources}}

\section{Introduction} \label{sec:intro} 

Studies of associated neutral hydrogen (\hi) 21\,cm absorption mainly focus on the interaction between active galactic nuclei (AGN) and the surrounding interstellar medium (ISM), as well as on the overall structure and kinematics of \hi\ gas and its physical properties and evolution across different types of AGNs (see \cite{2018A&ARv..26....4M} for a review). AGN are luminous cores of galaxies powered by accretion onto supermassive black holes (SMBHs), and their interaction with ISM plays a key role in star formation and galaxy evolution \citep{2008A&ARv..15..189S,2014dmp..book.....S}. The actual impact of AGN feedback depends on both AGN types and the properties of the ISM. Radio AGNs are typically hosted by early-type galaxies (ETGs), and observations of \hi\ allow us to investigate the connection between cold gas, star formation, and nuclear activity \citep[e.g.,][]{2001MNRAS.323..331M,2003A&A...404..861V,2006MNRAS.373..972G,2011MNRAS.418.1787C,2015A&A...575A..44G,2017A&A...604A..43M,2022MNRAS.516.4338D,2023ApJ...952..144Y,2024ApJ...973...48C}. In addition, radio AGNs show radio jets expanding through ISM, offering a unique opportunity to study how nuclear activity influences the ISM at different evolutionary stages \citep{2018MNRAS.479.5544M,2022A&A...659A.185M}.

As a powerful tracer, observation of the \hi\ 21 cm absorption depends only on the strength of background radio sources, which offers significant observational advantages. And in the large number of \hi\ absorption detections in radio sources, absorption typically occurs near the central region \citep{2018A&ARv..26....4M}, providing a valuable probe of ISM structure and dynamics in the vicinity of AGNs. The shape and width of the absorption profile can reveal whether ISM is undergoing orderly rotation or shows signs of irregular gas distributions \citep{1999ApJ...524..684G,2010A&A...513A..10S,2011A&A...535A..97M}, offering indirect insights into galaxy interactions. Recent studies \citep[e.g.,][]{2013Sci...341.1082M,2021A&A...647A..63S} have also shown that fast outflow of cold gas driven by AGN is a manifestation of AGN feedback. Observations of such outflows in \hi\ thus offer compelling evidence of the interaction between AGN energy output and ISM, helping us understand the physical mechanisms involved \citep{1999NewAR..43..509C,2013Sci...341.1082M,2018MNRAS.473...59A,2021A&A...647A..63S}. Also, recent studies \citep[e.g.,][]{2009A&A...505..559M,2014A&A...571A..67M,2016Natur.534..218T} have revealed evidence of cold gas (e.g., \hi\ and H$_2$) infalling toward the SMBH, providing clues for understanding the fueling of AGNs.

Observations and studies of \hi\ 21 cm absorption have been carried out for decades, with more than 200 absorbers detected \citep[e.g.,][]{2018A&ARv..26....4M,2021MNRAS.503.5385Z,2022MNRAS.516.2947S,Hu2023,2025ApJS..277...25H,2023ApJ...952..144Y,2024MNRAS.527.8511A,2025ApJS..276....6Z}. Early surveys, aiming to balance sensitivity and observation time limited by capabilities of available telescopes, primarily targeted bright sources ($S_{1.4\,\mathrm{GHz}} > 50\,\mathrm{mJy}$) \citep{1999ApJ...524..684G,2015A&A...575A..44G,2019MNRAS.482.5597A}, most of which are high-power radio systems ($\log(P_{\mathrm{1.4\,GHz}}/\mathrm{W\,Hz^{-1}}) > 24$). However, radio-faint sources, often found in ETGs, constitute the majority of the radio AGN population \citep{1997ApJ...479..642B,2005MNRAS.362....9B,2007MNRAS.381..211S,2012MNRAS.421.1569B,2019A&A...622A..17S,2022A&A...665A.144M}. And radio AGN activity shows a positive correlation with the stellar mass \citep[e.g.,][]{2012A&A...545A..15S,2019A&A...622A..17S}. \cite{2017A&A...604A..43M} extended the search for \hi\ absorption to relatively faint sources ($30\,\mathrm{mJy}< S_{1.4\,\mathrm{GHz}} < 50\,\mathrm{mJy}$) using Westerbork Synthesis Radio Telescope (WSRT), targeting a redshift range of $0.02 < z < 0.25$ and increasing the number of absorber detections.

To fill the absence in systematic observations of radio-faint sources ($S_{1.4\,\mathrm{GHz}} < 30~\mathrm{mJy}$) and study the low-power population, \cite{2023ApJ...952..144Y} performed pilot observations for a systematic \hi\ absorption survey of faint sources ($10\,\mathrm{mJy} < S_{1.4\,\mathrm{GHz}} < 30\,\mathrm{mJy}$) using the Five-hundred-meter Aperture Spherical radio Telescope (FAST). Their study found a lower detection rate of $\sim$11.5\% for sources with $\log(P_{1.4\,\mathrm{GHz}}/\mathrm{W\,Hz^{-1}}) = 21.8{-}23.7$ and an even lower detection rate of 6.7\% for sources with $\log(P_{1.4\,\mathrm{GHz}}/\mathrm{W\,Hz^{-1}}) < 23$. These newly detected absorbers show relatively narrow line widths and column densities consistent with previous studies, demonstrating the high sensitivity of FAST to detect weak absorbers with complex kinematics. However, due to the small number of pilot observations, the results of the detection rate and gas kinematics are limited to performing statistical comparison with those of high-power radio sources.

In this work, we complete the systematic survey of \hi\ absorption in low-power radio sources by combining new FAST observations and results from \cite{2023ApJ...952..144Y}. The final sample includes a total of 159 radio sources. The primary goal of this work is to use FAST to identify new \hi\ absorbers and to investigate the impact of radio power on the interaction between AGNs and ISM. This paper is organized as follows. In Section \ref{sec:obs}, we describe sample selection, FAST observations, and the data reduction process. Section \ref{sec:res} presents the observational results, including the detection rate of \hi\ absorbers and the properties of the detected absorbers. In Section \ref{sec:dis}, we classified the sample and discussed the results among different types of AGNs with a focus on how radio power influences the \hi\ content in host galaxies. In Section \ref{sec:sum}, we summarize the main conclusions. Throughout this work, we adopt a $\Lambda$CDM cosmology with $H_0 = 70\,\mathrm{km\,s^{-1}\,Mpc^{-1}}$, $\Omega_M = 0.3$, and $\Omega_\Lambda = 0.7$.

\section{Observations and Data Reduction} \label{sec:obs}

\subsection{Sample selection and Observations} \label{sec:sam}

\begin{figure*}[t!]
\includegraphics[width=\textwidth]{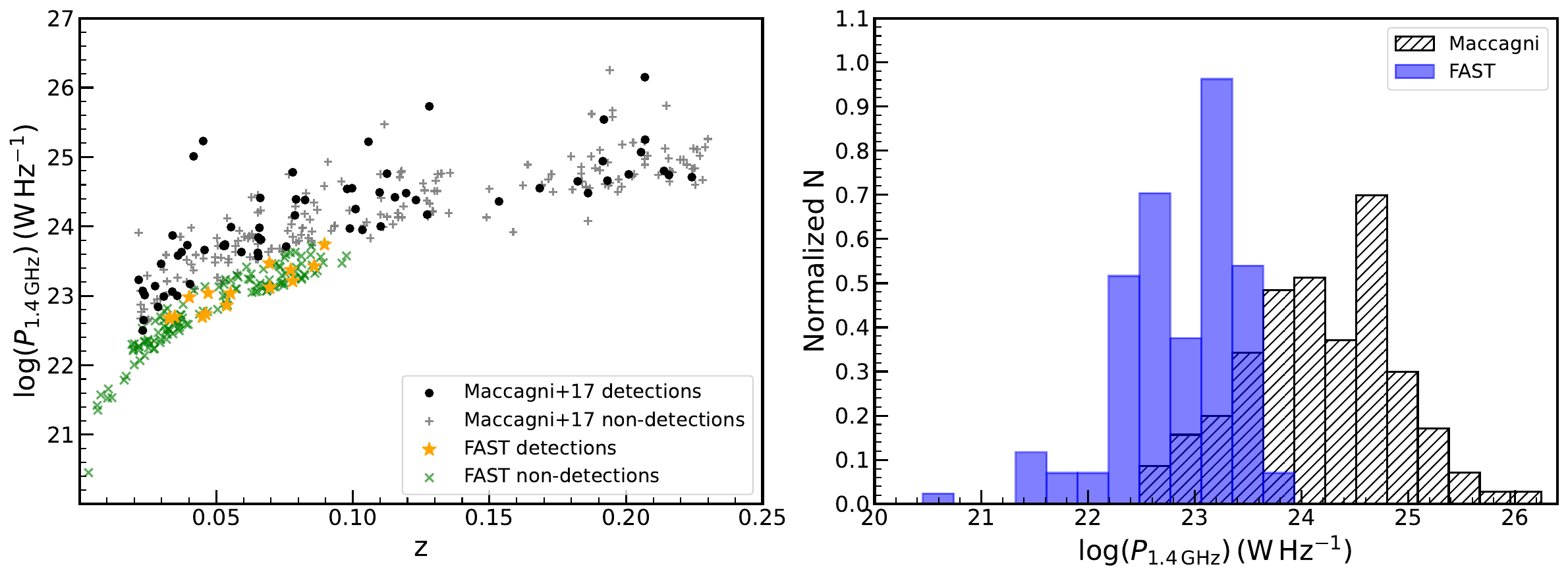}
\caption{Left: distribution of radio power vs. redshift for our sample and the sample of \cite{2017A&A...604A..43M}. Right: histograms of the radio power for our sample and the sample of \cite{2017A&A...604A..43M}.} 
\label{fig:1}
\end{figure*}

We constructed a low-power radio source sample to explore the less observed regime (Figure \ref{fig:1}) due to flux limits of previous studies \citep[e.g.,][]{2015A&A...575A..44G,2017A&A...604A..43M}. We selected our targets by crossmatching between the NRAO Very Large Array (VLA) Sky Survey (NVSS) catalog \citep{1998AJ....115.1693C} and the Sloan Digital Sky Survey Data Release 16 (SDSS DR16) galaxy catalog \citep{2020ApJS..249....3A}. Sources with 1.4 GHz radio flux densities in the range of 10 mJy $<S_{\rm 1.4 GHz}<$ 30 mJy were selected, while imposing a redshift of \( z < 0.1 \) to mitigate strong radio frequency interference (RFI). To avoid duplicated observations, these sources were also crossmatched with the second release of HI-MaNGA data \citep{2021MNRAS.503.1345S} and the Arecibo Legacy Fast ALFA survey catalogs \citep{2018ApJ...861...49H} to exclude objects with existing \hi\ 21 cm emission detections in signal-to-noise ratio \((\text{S/N}) > 5\). These excluded sources are gas-rich galaxies without \hi\ absorption  based on Arecibo observations. Applying these criteria, we compiled a sample of 159 low-power radio sources. 

Pilot observations of 29 sources were completed and analyzed by \cite{2023ApJ...952..144Y}, with subsequent observations of the remaining 130 sources conducted with the FAST PI program (PT2022\_0090, PI: Q. Yu) from 2022 September to 2023 April. Among the 130 sources, observations of two sources failed due to technical issues, resulting in 128 observed sources in our new observations. FAST has a beam size of $\sim$2.9$'$ and a channel width of 7.63 kHz (equal to a velocity resolution of 1.6 km s\(^{-1}\)) at 1.4 GHz, with pointing error less than or equal to 15$''$ \citep{Nan2006,Nan2011,Jiang20}. Observations were conducted in the ON-OFF mode and the OFF position is set to a nearby sky location to estimate background contributions.

\subsection{Characterization of the sample}

The radio sources in our sample are nearby galaxies (\(z < 0.1\)) with a radio power range of $\log(P_{1.4\,\mathrm{GHz}}/\mathrm{W\,Hz^{-1}}) = 20.5{-}23.7$. Following the method of \cite{2017A&A...604A..43M}, we find the majority of the radio galaxies (\(\sim 84\%\)) lie on the high-mass end of the red sequence (\(-24 < M_r < -21\), \(u-r \approx 2.65 \pm 0.60\)), representing typical massive early-type systems. The remaining radio galaxies exhibit bluer colors (\(u-r < 2\)) and are primarily composed of disk galaxies. Among the sample, nine radio sources exhibit clear signs of galaxy interactions in the DESI image, including ongoing mergers, prominent gaseous and stellar tidal tails, or double nuclei. These features indicate recent or ongoing interactions with companion galaxies. Most of these systems have also been classified as merging galaxies in previous studies (see Table \ref{tab:det}, Table \ref{tab:a1} and \ref{C} for details). Since these nine interacting sources are different from the rest of the sample, we treat them as a distinct subsample in our overall statistics, radio morphology classification, and Wide-field Infrared Survey Explorer (WISE) color analysis. In the rest sample, a small fraction of galaxies are classified as disk galaxies with the aforementioned method, while we notice there are uncertainties considering only photometric classification. In order to perform a direct comparison between our sample and previous studies, we keep the remaining sources and investigate the impacts of star-forming-dominated sources with multiband photometry and spectroscopy.

We classify the radio morphologies of our sources following the same criteria adopted by \cite{2014A&A...569A..35G,2015A&A...575A..44G} and \cite{2017A&A...604A..43M}. We classified the radio continuum morphology using the NVSS major-to-minor axis ratio and the peak-to-integrated flux ratio from the VLA Faint Images of the Sky at Twenty centimeters
(FIRST) survey \citep{Becker1995}. Compact sources typically have radio continuum emission confined within subgalactic scales (i.e., within a few kiloparsecs of the host), whereas extended sources have radio structures that span super-galactic scales \citep{2017A&A...604A..43M}. For a population of radio AGNs, the spatial extent of the radio continuum may be related to the age of nuclear activity. Extended (older) radio sources are less likely to interact directly with central gas regions of the galaxy, whereas compact sources are considered to be newly born (or restarted) radio AGNs \citep{1998PASP..110..493O,2003PASA...20...19M,2009AN....330..120F,2016AN....337....9O,2016AN....337..105S,2018A&ARv..26....4M}. These are usually still embedded in their host galaxies and strongly coupled with the ISM, potentially inducing strong disturbances in the gas (negative feedback). According to this scheme, the majority of sources in our sample are identified as extended sources ($\sim$76.9\%), while $\sim$17.0\% are classified as compact sources. The detailed classification is provided in Table \ref{tab:res}.

The WISE color-color diagram is an effective tool for classifying galaxies based on their mid-infrared (MIR) properties. Various studies employing MIR colors \citep[e.g.,][]{2010AJ....140.1868W,2012ApJ...753...30S,2013AJ....145....6J,2016MNRAS.462.2631M,2017A&A...604A..43M,2017MNRAS.465..997C,2020MNRAS.494.5161C,2025ApJS..277...25H} have utilized the all-sky data from WISE, which observed in four MIR bands: 3.4\,$\mu$m (W1), 4.6\,$\mu$m (W2), 12\,$\mu$m (W3), and 22\,$\mu$m (W4). The W3 band is sensitive to dust continuum \citep{2013ApJ...774...62L,2014ApJ...782...90C}. In dust-poor red-sequence galaxies, the 12\,$\mu$m luminosity is dominated by the old stellar population and is comparable to the 4.6\,$\mu$m luminosity. In contrast, galaxies rich in PAHs and dust show enhanced 12\,$\mu$m luminosity, increasing their W2 $-$ W3 colour. The W1 $-$ W2 colour is sensitive to heated dust.

In this study, we crossmatched sky coordinates of each source with the WISE catalog using VizieR catalog access tool \citep{2000A&AS..143...23O} and extracted WISE photometry. We use the MIR colors from the WISE to classify the dust content of the galaxies in our sample \citep{2010AJ....140.1868W}. Following the same criteria as adopted in \cite{2016MNRAS.462.2631M} and \cite{2017A&A...604A..43M}, we divide the sources into two categories: dust-poor sources ($\sim44.9\%$) and MIR-bright sources (12\,$\mu$m bright sources: $\sim16.3\%$; 4.6\,$\mu$m bright sources: $\sim31.3\%$). 

To characterize the AGN activity and star formation in our sample, we incorporate optical spectroscopic data from SDSS and utilize the Baldwin–Phillips–Terlevich (BPT) diagram \citep{1981PASP...93....5B} to classify these sources. We adopt the narrow-line BPT diagram using [O\,\textsc{iii}]/H$\beta$ versus [N\,\textsc{ii}]/H$\alpha$ \citep{1981PASP...93....5B,1987ApJS...63..295V,2006MNRAS.372..961K,2010MNRAS.403.1036C}, which separates galaxies into four distinct types of nuclear activity:

\begin{enumerate}
    \item \textit{Star forming.} Dominated by star formation;
    \item \textit{Seyfert.} AGNs powered by accretion onto SMBHs;
    \item \textit{Low-Ionization Nuclear Emitting Regions (LINERs ).} Weak AGNs with moderate nuclear activity, possibly driven by a mix of stellar processes and AGN processes \citep{2004ApJ...605..105C};
    \item \textit{Composites.} Transition objects with contributions from both star-formation and AGN activity.
\end{enumerate}

Based on the BPT diagram, a small fraction of the galaxies are identified as star-forming galaxies, comprising $\sim 21\%$ of the entire sample. As a comparison, the sample of \cite{2017A&A...604A..43M} also contains a fraction of star-forming galaxies based on the BPT classification (see Section \ref{bpt}). To make a direct comparison and investigate how star formation impacts the detection and properties of \hi\ absorbers, we therefore keep these star-forming-dominated sources in the following analysis.

\subsection{Data Reduction} \label{sec:dat}
 
The raw data were calibrated using the FAST spectral data reduction pipeline, HiFAST \citep{2024SCPMA..6759514J,2024RAA....24h5009L,2025RAA....25a5011X}. HiFAST is a dedicated pipeline designed for calibrating and imaging FAST observations, incorporating critical steps such as temperature calibration, baseline modeling, standing-wave mitigation, flux density calibration, RFI flagging, Doppler correction, gridding, and FITS cube generation. The corresponding \hi\ spectra were obtained through HiFAST. To optimize the S/N, the calibrated spectra were binned during postprocessing. The final rms noise levels span $\sim$0.13$-$0.62 mJy across the dataset, with a velocity resolution of $\sim$10$-$20 km/s. Considering the majority of the detected absorption profiles are relatively regular, we fitted these \hi\ profiles uniformly by the busy function \citep{2014MNRAS.438.1176W}. Parameters of the detected sources are tabulated in Table \ref{tab:res}, while nondetections exhibit a median rms noise of $\sim$0.24 mJy and a velocity resolution of 10 km/s, as detailed in Table \ref{tab:a1}. FAST achieves an average noise level $\sim$3.5 times lower than previous studies conducted with other telescopes at the same velocity resolution \citep{2015A&A...575A..44G,2017A&A...604A..43M}, which provides required sensitivity to reach comparable detection limits for fainter sources. For spectra in which emission and absorption features coexist, we adopt the same method as in \cite{2025ApJS..276....6Z}. For the two sources (SDSS J145631.33+243634.9 and SDSS J160151.51+024809.9) where the absorption line is blended with the emission line, we first mask the absorption feature and fit the emission component. We then subtract the fitted emission line from the original spectrum, after which the absorption line can be properly fitted.

\section{Results} \label{sec:res}    
Among 128 observed sources in our new observations, seven sources were severely affected by RFI and excluded from the following statistical analysis. Combining our results with those of \cite{2023ApJ...952..144Y} (26 sources), a total of 147 sources were included in the final statistical analysis.
We detected 12 new absorbers in our follow-up observations, and combined with the results of \cite{2023ApJ...952..144Y}, a total of 15 absorbers were identified in our sample. The detection rate is $\sim10.2^{+3.1}_{-2.0}\%$ in the radio power of $\log(P_{1.4\,\mathrm{GHz}}/\mathrm{W\,Hz^{-1}}) = 20.5-23.7$. In the left panel of Figure \ref{fig:1}, we show the distribution of radio power versus redshift for our sources compared with those of \cite{2017A&A...604A..43M}. At similar redshifts, our observations extend to lower radio powers ($\log(P_{1.4\,\mathrm{GHz}}/\mathrm{W\,Hz}^{-1}) < 24$), where the number of observed sources is increased by a factor of $\sim$2 compared to the sample of \cite{2017A&A...604A..43M}. The right panel of Figure \ref{fig:1} presents histograms of the two samples. The size of our sample is comparable to that of \cite{2017A&A...604A..43M}, both being of the same order of magnitude, but our sources are more strongly concentrated at lower radio powers.

In our sample, \hi\ emission lines are detected in 48 sources, yielding a detection rate of $\sim$32.7\% for \hi\ emission. Excluding interacting sources, the total detection rate of \hi\ gas in both absorption and emission is consistent with previous surveys of ETGs \citep{2012MNRAS.422.1835S}, which suggest $\sim$40\% of ETGs have detectable amounts of \hi.

\subsection{Physical Properties of the Absorbers}

The absorption line width, $W_{20}$, is measured at the 20\% peak flux density of each fitted profile, while the line center is defined at the midpoint of $W_{20}$. The \hi\ column density of the foreground gas can be calculated by integrating the observed optical depth profile over the velocity \citep{2018A&ARv..26....4M} ,

\begin{equation}
\nhi = 1.823 \times 10^{18}\ {T_s} \int \tau(v) {\rm d} v \text{ cm}^{-2},  
\label{eq:1}
\end{equation}

where $\tau(v)$ is the optical depth of the \hi\ absorber at Doppler-corrected velocity $v$ (kilometers per second), and $T_s$ is the spin temperature (kelvins). Under the assumption that $T_s \ll c_f T_c$, the optical depth $\tau(v)$ can be expressed as

\begin{equation}
\tau(v) = - \text{ln}\left(1 + \frac{\shi}{c_f\ S_{\text{1.4 GHz}}} \right) ,  
\label{eq:2}
\end{equation}

where $S_{\text{H\,\textsc{i}}}$ is the flux density of the H\,\textsc{i} absorption line, $S_{1.4\,\text{GHz}}$ is the continuum flux density at 1.4 GHz, and $c_f$ is the covering factor of the absorbing gas. In this study, we adopt $T_s = 100\,\text{K}$ and $c_f = 1$ for all sources, following the assumptions commonly used in previous works \citep[e.g.,][]{2015A&A...575A..44G,2017A&A...604A..43M,2020ApJ...900L..30C,2023ApJ...952..144Y}.

For the FAST sample, the optical depth (and upper limits) of the \hi\ absorption is derived using the FIRST 1.4\,GHz peak flux density. Given that \hi\ absorption in most sources predominantly occurs in the nuclear and central regions of the host galaxies (on scales of kpc or smaller), the beam of FIRST (\(\sim5^{\prime\prime}\)) more closely matches the spatial scale relevant for characterizing the core flux. In addition, since the majority of our sample sources exhibit extended radio morphologies, adopting the higher-resolution flux for the optical depth calculation is expected to be closer to their intrinsic values. The relevant measurements are listed in Table \ref{tab:res}.

However, caveats also should be considered for uncertainties due to unresolved observations and the missing flux issue from interferometric measurements. In our sample, the angular size of the sources varies across different redshifts, and the majority of the radio continuum sources are not well resolved even with FIRST observations, leading to uncertainties in estimating the central regions where absorption occurs. On the other hand, as a single-dish telescope, FAST does not suffer from missing flux, and the continuum flux densities are generally more consistent with those obtained from NVSS, which has lower angular resolution compared to FIRST. Therefore, we keep both continuum measurements for reference in Table \ref{tab:res} and Table \ref{tab:a1}, and careful considerations should be taken into account when comparing with different surveys.

A summary of the detections is provided in Table \ref{tab:res}, which lists the basic source properties. The peak optical depths ($\tau_{\rm peak}$) range from $\sim0.049$ to $\sim1.271$, and the integrated optical depths ($\tau_{\rm int}$) span $\sim3.13$ to $\sim62.48$ km s$^{-1}$. The absorbers have \hi\ column densities ($\nhi$) around $10^{20}$ to $10^{21}$ cm$^{-2}$. The absorption spectra with fitted profiles are shown in \ref{sec:B}. We treat lines with $\mathrm{S/N} > 4.5$ as detections and $3 < \mathrm{S/N} \leqslant 4.5$ as tentative detections. Four absorbers (SDSS J111700.10+323551.0, SDSS J124322.55+373858.0, SDSS J130404.99+075428.3, and SDSS J152659.44+355837.0) are regarded as tentative detections. 

Following the classification method of \cite{2017A&A...604A..43M}, we first examined optical images of the absorber sources. Several sources, based on their DESI images, appear to be undergoing mergers with spiral galaxies or exhibit extended gas and stellar tidal tails, suggesting recent or ongoing interactions with companion galaxies. In addition, we inspected DESI images and optical redshifts of nondetection sources. As a result, nine sources were classified as interacting systems, as listed in Table \ref{tab:res}.

\begin{deluxetable*}{lcccccccccccccr}
\setlength\tabcolsep{1pt}
\tablenum{1}
\tablecaption{Absorption line properties\label{tab:res}}
\tabletypesize{\scriptsize}
\tablewidth{\textwidth}
\tablehead{
\colhead{Source Name} & \colhead{$z$} & \colhead{$S_{\text{1.4\ GHz, NVSS}}$} & \colhead{$S_{\text{1.4\ GHz, FIRST}}$} & \colhead{log($P_{\text{1.4\ GHz}}$)} & \colhead{rms} & \colhead{$\Delta v$} & \colhead{$W_{20}$} & \colhead{$v_{\text{centroid}}$} & \colhead{$\tau_{\text{peak}}$} & \colhead{$\tau_{\text{int}}$} & \colhead{$N_{\rm HI}$} & \colhead{Radio type} & \colhead{WISE type} & \colhead{S/N}\\
\colhead{} & \colhead{} & \colhead{(mJy)} & \colhead{(mJy $\text{beam}^{-1}$)} &
\colhead{(W $\text{Hz}^{-1}$)} & \colhead{(mJy)} & \colhead{($\text{km s}^{-1}$)} & \colhead{($\text{km s}^{-1}$)} &
\colhead{($\text{km s}^{-1}$)} & \colhead{} & \colhead{($\text{km s}^{-1}$)} & \colhead{($10^{20}\ \text{cm}^{-2}$)} &
\colhead{} & \colhead{} & \colhead{}
}
\decimalcolnumbers
\startdata
SDSS J110852.61+510225.7 & 0.06964 & 10.8 & 9.7 & 23.12 & 0.24 & 10 & 69.0 & 101.6 & 0.135 $\pm$ 0.022 & 6.09 $\pm$ 0.87 & 9.92 $\pm$ 1.41 & E & 4.6\,$\mu$m & 5.1 \\
SDSS J111700.10+323551.0 & 0.03479 & 17.5 & 17.6 & 22.70 & 0.29 & 15 & 77.1 & 1.2 & 0.059 $\pm$ 0.017 & 3.13 $\pm$ 0.70 & 5.73 $\pm$ 1.28 & C & dp & 3.5 \\
SDSS J113402.18+263224.5 & 0.07724 & 15.6 & 8.8 & 23.38 & 0.29 & 10 & 172.9 & 62.8 & 0.311 $\pm$ 0.038 & 14.70 $\pm$ 1.57 & 15.06 $\pm$ 1.61 & E & dp & 8.0 \\
SDSS J124322.55+373858.0 & 0.08584 & 14.1 & 10.8 & 23.43 & 0.22 & 20 & 122.6 & $-$274.6 & 0.093 $\pm$ 0.022 & 6.30 $\pm$ 1.61 & 8.83 $\pm$ 2.25 & I & 4.6\,$\mu$m & 4.5 \\
SDSS J130404.99+075428.3 & 0.04614 & 10.7 & 9.3 & 22.74 & 0.13 & 20 & 239.1 & 96.8 & 0.049 $\pm$ 0.014 & 6.21 $\pm$ 1.08 & 9.83 $\pm$ 1.71 & C & dp & 3.3 \\
SDSS J133159.83+015404.1 & 0.07797 & 10.5 & 8.5 & 23.21 & 0.13 & 20 & 160.0 & 64.7 & 0.078 $\pm$ 0.020 & 7.30 $\pm$ 1.01 & 10.77 $\pm$ 1.50 & E & 4.6\,$\mu$m & 4.8 \\
SDSS J134159.72+294653.5 & 0.04490 & 10.3 & 9.6 & 22.70 & 0.14 & 10 & 34.7 & 88.3 & 0.316 $\pm$ 0.020 & 6.52 $\pm$ 0.36 & 11.07 $\pm$ 0.61 & E & dp & 18.6 \\
SDSS J141803.26+272800.5 & 0.06946 & 24.3 & 21.0 & 23.47 & 0.33 & 15 & 427.9 & $-$62.8 & 0.098 $\pm$ 0.007 & 27.30 $\pm$ 1.56 & 43.01 $\pm$ 2.46 & I & 4.6\,$\mu$m & 6.0 \\
SDSS J142633.10+165047.8 & 0.05379 & 10.3 & 7.8 & 22.86 & 0.17 & 10 & 187.4 & 55.0 & 0.162 $\pm$ 0.014 & 16.58 $\pm$ 1.29 & 22.83 $\pm$ 1.77 & C & 12\,$\mu$m & 7.1 \\
SDSS J145631.33+243634.9 & 0.03280 & 19.1 & 20.5 & 22.68 & 0.44 & 10 & 40.5 & 48.7 & 0.353 $\pm$ 0.032 & 9.11 $\pm$ 0.50 & 17.80 $\pm$ 0.97 & C & 4.6\,$\mu$m & 13.9 \\
SDSS J152659.44+355837.0 & 0.05521 & 14.5 & 12.6 & 23.03 & 0.32 & 10 & 216.1 & 187.8 & 0.122 $\pm$ 0.168 & 15.54 $\pm$ 2.18 & 24.68 $\pm$ 3.46 & I & 4.6\,$\mu$m & 4.5 \\
SDSS J160151.51+024809.9 & 0.03291 & 18.5 & 20.6 & 22.67 & 0.46 & 10 & 122.6 & $-$1.8 & 0.575 $\pm$ 0.039 & 19.70 $\pm$ 0.82 & 39.93 $\pm$ 1.67 & E & 4.6\,$\mu$m & 19.7 \\
SDSS J133245.62+263449.3\tablenotemark{a} & 0.04701 & 21.1 & 20.4 & 23.04 & 0.31 & 10 & 106.9 & $-$26.8 & 0.095 $\pm$ 0.018 & 5.08 $\pm$ 0.69 & 8.95 $\pm$ 1.22 & C & 4.6\,$\mu$m & 5.9 \\
SDSS J151205.57+305023.3\tablenotemark{a} & 0.08963 & 27.8 & 3.9 & 23.74 & 0.40 & 10 & 172.7 & 93.0 & 1.271 $\pm$ 0.427 & 62.48 $\pm$ 5.74 & 16.06 $\pm$ 1.48 & E & dp & 7.1 \\
SDSS J162846.13+252940.9\tablenotemark{a} & 0.04010 & 25.2 & 27.3 & 22.98 & 0.38 & 20 & 131.6 & $-$195.9 & 0.066 $\pm$ 0.016 & 5.96 $\pm$ 0.89 & 11.78 $\pm$ 1.77 & E & dp & \nodata
\enddata
\tablecomments{
Column (1): the source name. Column (2): redshift from SDSS spectroscopy \citep{2020ApJS..249....3A}.
Column (3): radio continuum flux at 1.4 GHz from NVSS \citep{1998AJ....115.1693C}.
Column (4): radio continuum peak flux at 1.4 GHz from FIRST \citep{2015ApJ...801...26H}.
Column (5): radio power at 1.4 GHz.
Column (6): 1$\sigma$ noise measured at re-binned velocity resolution.
Column (7): re-binned velocity resolution.
Column (8): line width measured at 20\% of the peak flux.
Column (9): line centroid of the absorption.
Column (10): peak optical depth.
Column (11): integrated optical depth.
Column (12): estimated \hi\ column density assuming $T_s=100$ K and $c_f=1$.
Column (13): radio morphological type.
Column (14): WISE classification type (dp refers to dust-poor sources).
Column (15): the peak signal-to-noise ratio defined as $f_{\text{peak}}/rms$.
}
\tablenotetext{a}{The detailed data and measurements of this source are adopted from \cite{2023ApJ...952..144Y}.}
\end{deluxetable*}

\subsection{Special Absorbers}

In our sample of low-power radio sources, most absorbers show relatively narrow absorption profiles, and their line centers are near the systemic velocity. However, a small fraction shows significantly redshifted or blueshifted features. Among new detections, four absorbers (SDSS J110852.61+510225.7, SDSS J124322.55+373858.0, SDSS J152659.44+355837.0, and SDSS J162846.13+252940.9) exhibit redshifted or blueshifted velocities relative to host galaxies with $|v_{\text{centroid}}| > 100~\mathrm{km~s^{-1}}$, indicating candidates of gas inflows or outflows. Spatially resolved observations are needed to further investigate their origins.

Previous studies \citep{2015A&A...575A..44G,2017A&A...604A..43M} have shown that \hi\ absorption profiles exhibit a wide range of widths, shapes, and kinematic properties, but rarely report cases where absorption and associated emission lines coexist in the same profile. This peculiar line profile has been discovered in recent surveys \citep[e.g.,][]{2025ApJS..276....6Z}. In our sample, two such absorbers, SDSS J145631.33+243634.9 and SDSS J160151.51+024809.9, show absorption lines near the center of emission components. These emission features may originate from nearby interacting galaxies or from the same host galaxy as the absorber. In the \hi\ spectra of these two absorbers, the emission lines exhibit a symmetric double-peaked profile, which is likely associated with \hi\ gas in rotating disks. The coexistence of absorption and emission is rare and requires high spatial resolution observations, such as very long baseline interferometry (VLBI), to resolve host structures and understand mechanisms behind their formation.

\section{Discussion}\label{sec:dis}

 \subsection{Detection rate of Absorbers in low-power radio sources}\label{sec:det}

Combining our results with those of \cite{2023ApJ...952..144Y}, we find that, excluding sources affected by RFI, \hi\ absorption is detected in 15 out of a total of 147 sources, yielding a detection rate of $\sim10.2^{+3.1}_{-2.0}\%$. The $1\sigma$ error bars are estimated using the binomial statistics method of \cite{2011PASA...28..128C}. Taking into account the uncertainties, the detection rate for low-power radio sources ($\log(P_{1.4\,\mathrm{GHz}}/\mathrm{W\,Hz}^{-1}) < 24$) appears relatively lower than that in previous targeted surveys \citep[e.g.,][]{1989AJ.....97..708V,2010MNRAS.406..987E,2011MNRAS.418.1787C,2015A&A...575A..44G,2017A&A...604A..43M}. The upper limits of peak optical depth ($\tau_{\rm peak}$) for nondetections in our sample are estimated using a $3\sigma$ rms at $\sim$10 km\,s$^{-1}$ resolution. These upper limits are listed in Table \ref{tab:a1}. The median $\tau_{\rm peak}$ is $\sim$0.05, comparable to WSRT observations \citep[median $\tau_{\rm peak} \sim$0.04;][]{2017A&A...604A..43M}. Thus, the survey depth of our FAST observations is similar to WSRT, allowing a fair comparison of detection rates in overlapping regions of radio power and redshift. 

In previous studies \citep{2017A&A...604A..43M,2024ApJ...973...48C}, the \hi\ absorption detection rate shows no significant dependence on radio power. However, with our enlarged sample in the low radio power regime, we find that low-power radio sources exhibit a relatively lower \hi\ absorption detection rate compared to their higher-power counterparts at similar redshifts. 

\begin{figure}[t!]
\includegraphics[width=0.45\textwidth]{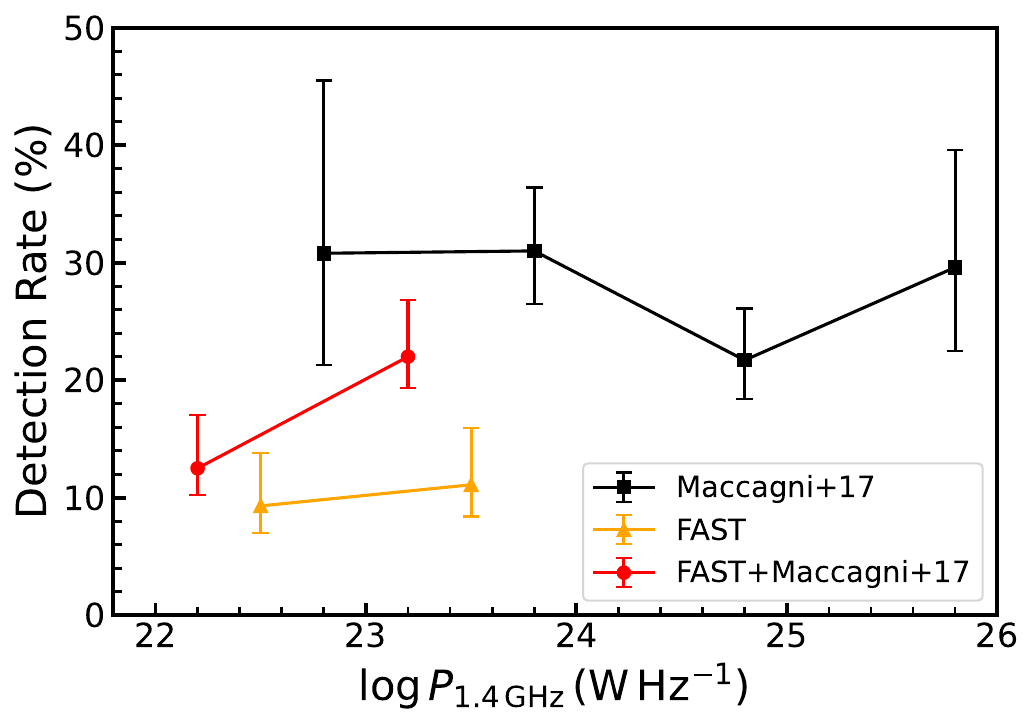}
 \caption{Comparison of the detection rate between our sample and that of \cite{2017A&A...604A..43M}, and the overall detection rates of the two samples. The detection rates of each sample are divided into four bins based on the radio power. The four bins are (1) $\text{log}(P_{\text{1.4 GHz}}/\text{W Hz}^{-1})<23$; (2) $23\leqslant \text{log}(P_{\text{1.4 GHz}}/\text{W Hz}^{-1})<24$; (3) $24\leqslant \text{log}(P_{\text{1.4 GHz}}/\text{W Hz}^{-1})<25$; (4) $\text{log}(P_{\text{1.4 GHz}}/\text{W Hz}^{-1})\geqslant25$.} 
\label{fig:2}
\end{figure}

To investigate the dependence of \hi\ absorption detection rate on radio power, in Figure \ref{fig:2} we plot the detection rates of our sample and that of \cite{2017A&A...604A..43M} as a function of radio power. The sources are divided into four power bins: $\text{log}(P_{\text{1.4 GHz}}/\text{W Hz}^{-1})<23$, $23\leqslant \text{log}(P_{\text{1.4 GHz}}/\text{W Hz}^{-1})<24$, $24\leqslant \text{log}(P_{\text{1.4 GHz}}/\text{W Hz}^{-1})<25$ and $\text{log}(P_{\text{1.4 GHz}}/\text{W Hz}^{-1})\geqslant25$. As shown in Figure \ref{fig:2}, within the uncertainties, detection rates in different power bins are similar across both samples. In the sample of \cite{2017A&A...604A..43M}, the detection rate of sources with $\log(P_{1.4\,\mathrm{GHz}}/\mathrm{W\,Hz^{-1}}) < 24$ is about 3 times higher than that of our sample ($\sim10.2^{+3.3}_{-2.1}\%$). Moreover, Figure \ref{fig:2} shows that when combining our sources with those of \cite{2017A&A...604A..43M}, the detection rate in the $\log(P_{1.4\,\mathrm{GHz}}/\mathrm{W\,Hz^{-1}}) < 23$ bin is lower than that in the $23 \leqslant \log(P_{1.4\,\mathrm{GHz}}/\mathrm{W\,Hz^{-1}}) < 24$ bin. These results suggest that low-power sources with $\log(P_{1.4\,\mathrm{GHz}}/\mathrm{W\,Hz^{-1}}) < 24$ may have relatively low \hi\ absorption detection rates, while the rate increases with radio power and then remains roughly constant.

Several factors may contribute to the relatively low detection rate in our sample. This may reflect a change in gas conditions under low radio power; low-power radio sources may reside in poor \hi\ environments or have less active SMBHs. Another possible explanation is the dilution by \hi\ emission \citep{2014MNRAS.440..696A,2018MNRAS.476.3580C}. At low redshifts (z $<$ 0.1), \hi\ emission is more easily detectable, and some \hi\ absorption in our sample might be diluted by the emission component (48 sources, $\sim32.7\%$ in our sample detected \hi\ emission line).

Additionally, the radio morphology of sources significantly affects the detection rate. Following the classification methods adopted by \cite{2014A&A...569A..35G,2015A&A...575A..44G} and \cite{2017A&A...604A..43M}, we classified the sample. Extended sources dominate in our low-power radio sample. In contrast, the WSRT sample \citep{2017A&A...604A..43M} contains a more balanced fraction of extended and compact sources. As shown in Table \ref{tab:det}, the detection rate for compact sources is $20.0^{+10.1}_{-5.7}\%$, significantly higher than that of extended sources, which is $6.2^{+3.1}_{-1.6}\%$. Although both rates are slightly lower than those in higher-power samples, the trend is consistent with that observed by \cite{2017A&A...604A..43M}: compact sources exhibit a higher detection rate. Therefore, the dominance of extended sources in our sample may contribute to the overall lower detection rate, and this may also suggest that low-power sources are dominated by extended sources. 

Moreover, as shown in Table \ref{tab:det}, interacting sources also show a higher detection rate ($33.3^{+17.5}_{-11.3}\%$) compared to noninteracting sources, though still lower than the detection rate among interacting sources in the WSRT sample \citep{2017A&A...604A..43M}.

\begin{deluxetable}{lccc}\setlength\tabcolsep{1.0pt}
\tablenum{2}
\tablecaption{Statistics of the sample\label{tab:det}}
\tabletypesize{\footnotesize}
\tablewidth{\textwidth}
\tablehead{
\colhead{Classification} & \colhead{Number of sources} & \colhead{Detections} & \colhead{Detection rate (\%)}
}
\decimalcolnumbers
\startdata
All sources & 147 & 15 & $10.2^{+3.1}_{-2.0}\%$ \\
\hline
Radio morphology classification\\
Extended sources & 113 & 7 & $6.2^{+3.1}_{-1.6}\%$ \\
Compact sources & 25 & 5 & $20.0^{+10.1}_{-5.7}\%$ \\
\hline
WISE colour classification\\
Dust-poor sources & 66 & 6 & $9.1^{+4.8}_{-2.4}\%$ \\
12\,$\mu$m bright sources & 24 & 1 & $4.2^{+8.4}_{-1.3}\%$ \\
4.6\,$\mu$m bright sources & 46 & 5 & $10.9^{+6.3}_{-3.1}\%$\\
\hline
Interacting sources & 9 & 3 & $33.3^{+17.5}_{-11.3}\%$
\enddata
 \tablecomments{The columns are (1) the classification method of the sources; (2) the number of observed source; (3) the number of \hi\ absorption detections; (4) the detection rates.}

\end{deluxetable}

\subsection{Kinematics and the Origin of the Absorbing Gas}\label{sec:kin}

\begin{figure*}[t!]
\includegraphics[width=\textwidth]{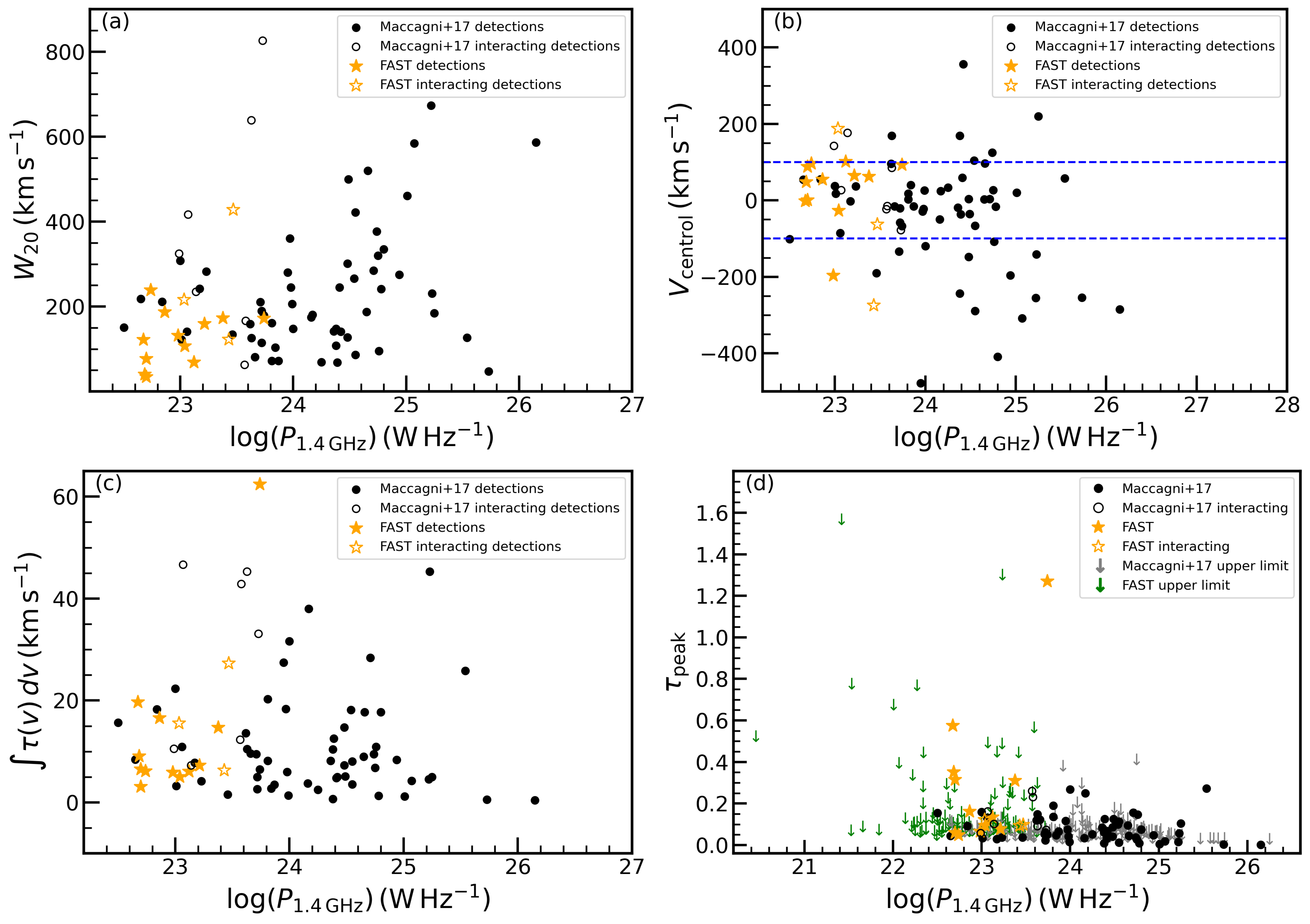}
\caption{(a) Full width measured at 20\% of the intensity ($W_{20}$) of the \hi\ profiles vs. the radio power of the sources. (b) Line centroid offset with respect to the systemic velocity vs. the radio power of the sources. The fine dashed lines in blue show the interval of $\pm 100\ \text{km s}^{-1}$. (c) Integrated optical depth vs. the radio power of the sources. (d) Peak optical depth vs. the radio power of the sources.} 
\label{fig:16}
\end{figure*}

Associated \hi\ absorption in radio AGNs provides insights into the origin and physical conditions of the cold gas. High spatial resolution \hi\ studies have revealed diverse absorbing structures \citep{2004MNRAS.352...49B,2010A&A...513A..10S,2010A&A...515A..67S,2013MNRAS.435L..58M,2014A&A...571A..67M}：including (circumnuclear) disks \citep{1999ApJ...524..684G,2008A&A...485L...5M,2010A&A...513A..10S}, high-velocity clouds \citep[HVCs;][]{1999NewAR..43..509C,2012A&A...546A..22S}, outflows/kinematically disturbed gas \citep{2013Sci...341.1082M,2021A&A...647A..63S}, and gas infalling toward the SMBH \citep[][]{2009A&A...505..559M,2014A&A...571A..67M,2016Natur.534..218T}.

We compared the morphological and kinematic characteristics of associated \hi\ absorption in sources with different radio power ranges. We compared our sample with that of \cite{2017A&A...604A..43M}. Figures \ref{fig:16}(a) and (b) show relationships between \hi\ absorption line width at 20\% of peak flux ($W_{20}$) and the 1.4 GHz radio power, and between the velocity offset of absorption line centroid from the systemic velocity ($\Delta v$) and $P_{1.4\,\mathrm{GHz}}$, respectively. Following the criteria adopted by \cite{2017A&A...604A..43M} and \cite{2023ApJ...952..144Y}, we classify absorption lines with $W_{20} > 400~\mathrm{km~s^{-1}}$ as broad. Taking into account galaxy rotation velocities and uncertainties in optical redshift measurements, and to maintain consistency with previous studies \citep{2014A&A...569A..35G,2015A&A...575A..44G,2017A&A...604A..43M, 2023ApJ...952..144Y}, we classify a detection as blueshifted/redshifted if the velocity offset of the line is larger than $\pm100~\mathrm{km\,s^{-1}}$.

As shown in Figures \ref{fig:16}(a) and (b), most of the \hi\ absorbers detected by FAST exhibit relatively narrow line widths and are centered near the systemic velocity. The only source in our sample with a broad line ($W_{20} > 400~\mathrm{km~s^{-1}}$) is SDSS J141803.26+272800.5, whose host galaxy is a merger system (see Table \ref{tab:res}). This result is consistent with previous findings \citep{2017A&A...604A..43M}, at low radio powers ($\log P_{1.4\,\mathrm{GHz}}/\mathrm{W\,Hz^{-1}} < 24$), most \hi\ absorption profiles are narrow and centered near the systemic velocity, and broad lines are only detected in merging systems. This suggests that the dominant \hi\ absorption component in low-power radio AGNs is likely associated with regularly rotating structures. And ETGs can also host \hi\ disks \citep{2010MNRAS.409..500O,2012MNRAS.422.1835S}. For absorbers detected against nuclear radio components, narrow widths ($< 100\ \text{km s}^{-1}$) near systemic velocity may originate from large-scale disks (e.g., dust lanes), while broader profiles (several hundred kilometers per second) may arise from circumnuclear disks, like Cygnus A \citep{1995ApJ...449L.131C,2010A&A...513A..10S} and Centaurus A \citep{1983ApJ...264L..37V,2010A&A...515A..67S,2012A&A...546A..22S}. Previous studies have shown that \hi\ 21\,cm absorption arising from rotating disks typically exhibits symmetric profiles with line widths of a few hundred kilometers per second and centroids close to the systemic velocity \citep{1999ApJ...524..684G,2021A&A...654A..94M}. Therefore, \hi\ absorption in low-power AGNs may predominantly originate from stable, rotating gas disks (like large-scale disks) in host galaxies.

Absorption features offset from systemic velocity or with broad redshift/blueshift wings can trace infalling gas or outflows, as observed in NGC 315 \citep{2009A&A...505..559M}, PKS B1718--649 \citep{2014A&A...571A..67M}, B2 1504+377 \citep{2008MNRAS.384L...6K}, NGC 1266 \citep{2011ApJ...735...88A}, 4C 12.50 \citep{2013MNRAS.435L..58M}, and PKS B1740--517 \citep{2015MNRAS.453.1249A}. As shown in Figure \ref{fig:16}(b), although most absorbers lie near the systemic velocity, a few show noticeable velocity offsets, indicating that both infalling and outflow gas are still present in low-power radio AGNs. For instance, the \hi\ absorption profile of SDSS J110852.61+510225.7 exhibits a slight redshift with a relatively narrow and symmetric profile (see Table \ref{tab:res} and \ref{sec:B}). The source SDSS J110852.61+510225.7 is classified as Seyfert-type AGN, with a spectroscopic redshift of \(z = 0.06964 \pm 0.00002\) \citep{2020ApJS..249....3A}. It is also identified as an extended source and a 4.6\,µm bright system. So such a narrow redshifted feature may originate from infalling gas clouds toward the SMBH \citep{2009A&A...505..559M,2014A&A...571A..67M}. However, given that the velocity shift is not significant, the \hi\ absorber may also originate from a rotating \hi\ disk. SDSS J152659.44+355837.0 shows a more prominent and asymmetric redshifted component, with a line width of $\sim216.1~\mathrm{km\,s^{-1}}$ (see Table \ref{tab:res} and \ref{sec:B}). This redshifted asymmetry could also arise from infalling gas clouds toward the SMBH \citep{2009A&A...505..559M,2014A&A...571A..67M}, though it may alternatively be explained by partial alignment of absorbing gas with AGN jets \citep{2001ApJ...554L.147P,2010A&A...515A..67S,2021A&A...654A..94M}. However, since it is an interacting source, the blueshift of the \hi\ absorption line may simply reflect unsettled gas resulting from the ongoing interaction. In addition, HVCs along the line of sight can also manifest as narrow \hi\ absorption features \citep{1999NewAR..43..509C,2010AJ....139...17A}. For instance, the origin of the redshifted narrow \hi\ component observed in 4C 31.04 remains uncertain \citep{2012A&A...546A..22S,2024A&A...688A..84M}, and one possible explanation is the presence of HVCs. SDSS J124322.55+373858.0, classified as an interacting system (Table \ref{tab:res} and \ref{sec:B}), displays a narrow and significantly blueshifted \hi\ absorption profile, with a velocity offset of $-274.6~\mathrm{km\,s^{-1}}$ from the systemic velocity. It is also classified as Seyfert-type AGN. Such blueshifted features may be driven by jet-accelerated outflows \citep{2013Sci...341.1082M,2021A&A...647A..63S}. Compared with high-power radio sources ($\log P_{1.4\,\mathrm{GHz}}/\mathrm{W\,Hz^{-1}} > 24$), such large blueshifted absorptions are rarer in the low-power regime. Assuming absorbing gas is associated with outflow, we estimate an upper limit on \hi\ mass outflow rate following \citep{2002ASPC..254..292H}
\begin{equation}
    \dot{M}_{\hi}\sim 30\frac{r_{\star}}{\rm kpc}\frac{N_{\hi}}{\rm 10^{21}\ cm^{-2}}\frac{v}{\rm 300\ km\ s^{-1}}\frac{\Omega}{4\pi}\ M_{\odot}\ {\rm yr^{-1}},
\end{equation}

\begin{figure*}[ht!]
\includegraphics[width=\textwidth]{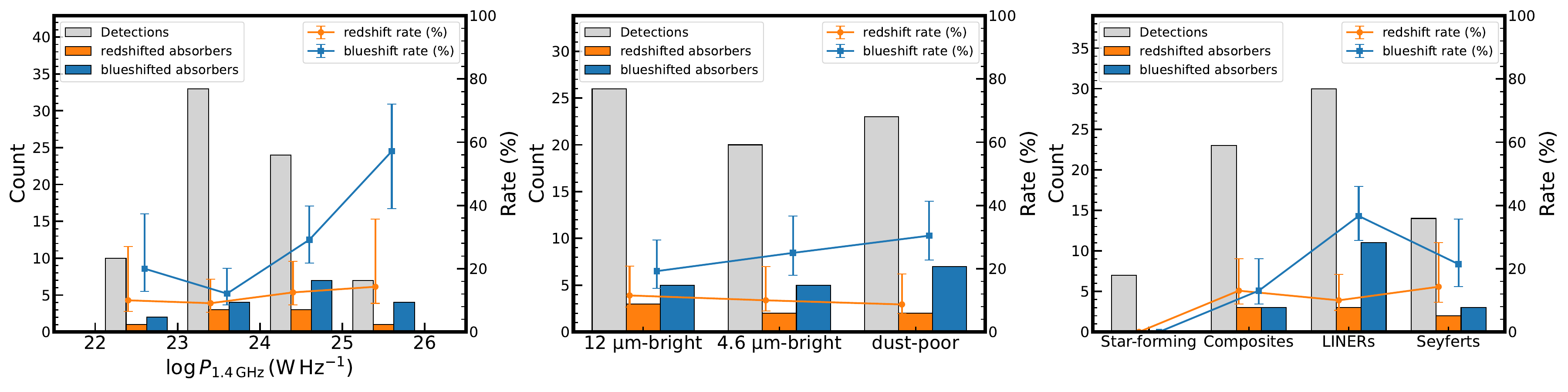} 
\caption{The number of detected absorbers, redshifted absorbers and blueshifted absorbers in our sample, and the WSRT sample \citep{2017A&A...604A..43M}, together with the fractions of redshift/blueshift rates among the detected absorbers. Left: as a function of radio power bins; Middle: according to the WISE color-color classification; Right: according to the BPT diagram classification.
} 
\label{fig:flow}
\end{figure*}

where $r_{\star}$ is the radius of outflow region, $N_{\hi}$ is the column density assuming $T_s=100$ K and covering factor $c_f=1$, $v$ is outflow velocity, and $\Omega$ is the solid angle subtended by outflow, assumed to be $\pi$. As neither FAST nor VLA resolves the continuum source, we adopt a typical upper-limit radius of $r_{\star}=1$~kpc, considering that most \hi\ outflows are confined to central kiloparsec scales \citep{2018A&ARv..26....4M}. With $N_{\hi} = 8.83\times10^{20}~\mathrm{cm^{-2}}$ and $v = 274.6 ~\mathrm{km\,s^{-1}}$, the estimated outflow rate is $\sim$6.0~$M_{\odot}\,\mathrm{yr^{-1}}$. This calculation provides only an estimate of the upper limit of the \hi\ outflow mass and does not represent a direct measurement of the true physical scale of the outflow. Significant uncertainties are associated with this estimate, and future observations with higher spatial resolution will be required to reliably constrain the properties and impact of potential outflows. Given the large velocity offset of absorption centroid and its narrow width ($W_{20} \sim 122.6~\mathrm{km\,s^{-1}}$), we speculate that absorbing structure may be located at a relatively large distance from AGN nucleus. Otherwise, tidal forces near the SMBH would likely broaden the line and shift the velocity centroid \citep{2012A&A...546A..22S}. This suggests that high velocity clouds (HVCs) located farther from nucleus may contribute to the observed absorption. Similarly, SDSS J162846.13+252940.9 exhibits a narrow, blueshifted profile, and both AGN-driven outflows and HVCs are possible origins \citep{2023ApJ...952..144Y}. Due to the limited spatial resolution of FAST, follow-up high-resolution interferometric \hi\ observations (e.g., VLBI) are essential for resolving the location and kinematics of absorbing gas, and for investigating possible jet-ISM interactions.

For absorbers with $v > 100~\mathrm{km\,s^{-1}}$, we classify them as redshifted absorbers, while those with $v < -100~\mathrm{km\,s^{-1}}$ are regarded as blueshifted absorbers. Redshifted or blueshifted \hi\ absorption lines may arise from \hi\ gas participating in inflow or outflow processes. Since the sources exhibiting such velocity shifts in both samples are all classified as AGN based on the BPT diagram, it is unlikely that these redshifted/blueshifted features are predominantly produced by rotating gas disks associated with star formation; instead, they are more plausibly related to AGN activity. We therefore tend to regard the sources showing redshifted or blueshifted absorption as inflow/outflow candidates. However, because the velocity offsets relative to the systemic velocity are not particularly large (e.g., $|v| > 1000~\mathrm{km\,s^{-1}}$) and still fall within the typical velocity range of rotating disks, the origin in rotating disks cannot be entirely ruled out. In the left panel of Figure \ref{fig:flow}, we present the total numbers of detected absorbers, redshifted absorbers, blueshifted absorbers, and fractions of redshifted/blueshifted absorbers across different radio power bins, combining results from our sample with those of the WSRT sample \citep{2017A&A...604A..43M}. By comparing detected \hi\ absorbers across different radio power ranges, we find that the fraction of \hi\ blueshifted absorbers increases with radio power, while the fraction of redshifted absorbers remains constant. These results may indicate the effect of radio emission of AGN on driving atomic gas outflows. As shown in the right panel of Figure \ref{fig:flow}, all redshifted/blueshifted absorbers in our sample are classified as AGN, with Seyfert and LINER systems dominating both samples. We tend to regard these sources as inflow/outflow candidates. This indicates the connection between gas participating in inflow/outflow processes and AGN activity, consistent with theoretical expectations and simulations in which energetically active AGN drive gas outflows \citep{2014MNRAS.444.2355C,2018NatAs...2..198H}.

In Figure \ref{fig:16}(c) and (d), we compare the integrated and peak optical depths of our sample with those from \cite{2017A&A...604A..43M}. For the integrated optical depth, our detections are broadly consistent with the distribution found in \cite{2017A&A...604A..43M}. Previous studies have shown that the peak optical depth of \hi\ absorption typically ranges from $\tau_{\rm peak} \sim 0.008$ to $\sim 0.15$ \citep{2017A&A...604A..43M,2024arXiv241206988M}, and most of the sources detected in our FAST observations fall within this range as well. Detections with low optical depth are relatively rare, likely due to the sensitivity limitations of current observations. However, in Figure \ref{fig:16}(d), we identify five sources with significantly higher peak optical depths ($\tau_{\rm peak} > 0.3$), which also exhibit high column densities $N_{\hi}$ (see Table \ref{tab:res}). Such high optical depths are uncommon and may indicate that these sources lie particularly dense regions of the surrounding environment, resulting in enhanced \hi\ absorption.

\subsection{Radio morphology Classification}\label{sec:uncertain}

\begin{figure*}[ht!]
\includegraphics[width=\textwidth]{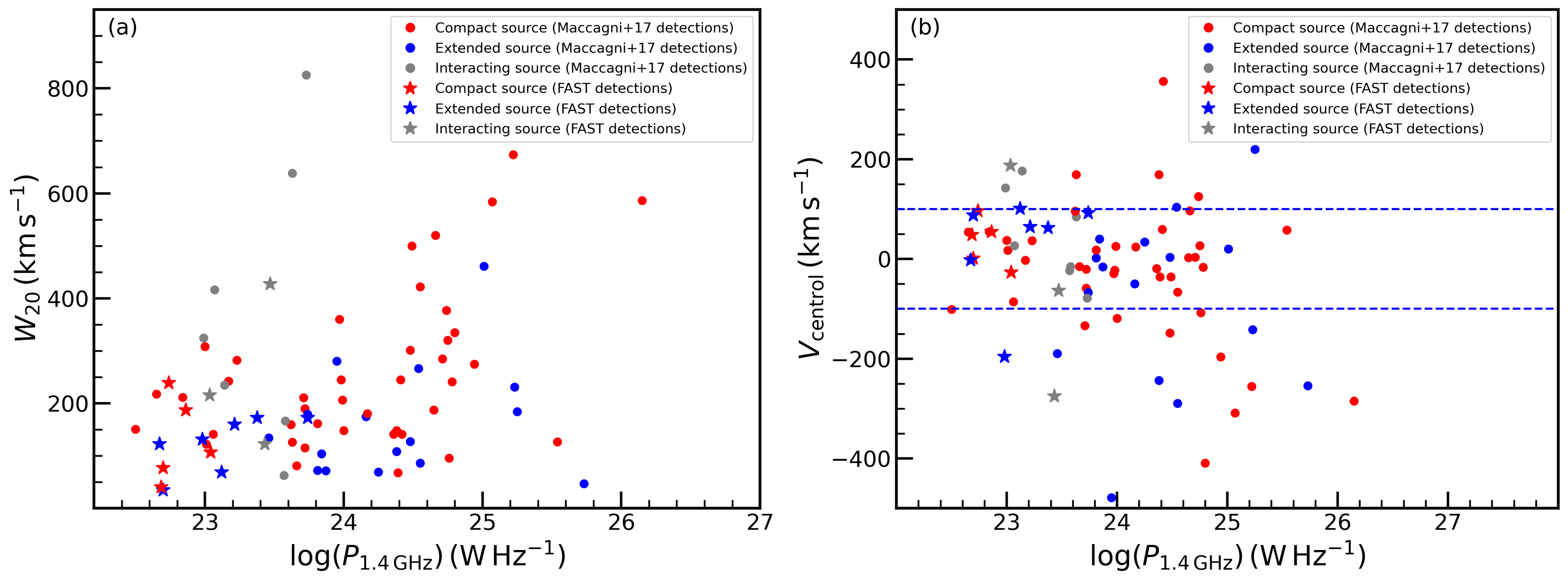}
\caption{Sources are classified according to the extension of their radio continuum. (a) Full width measured at 20\% of the intensity ($W_{20}$) of the \hi\ profiles vs. the radio power of the sources. (b) Line centroid offset with respect to the systemic velocity vs. the radio power of the sources. The fine dashed lines in blue show the interval of $\pm 100\ \text{km s}^{-1}$.} 
\label{fig:17}
\end{figure*}

Since \hi\ absorption traces foreground gas along the line of sight toward radio continuum sources, the morphology and structure of radio continuum can also affect observed absorption profile \citep{2021A&A...654A..94M}. We classified the noninteracting sources in our sample into compact and extended radio sources based on their radio morphology (see Tables \ref{tab:res}, \ref{tab:det} and \ref{tab:a1} for details). \hi\ absorbers detected in gigahertz peaked spectrum (GPS)/ compact steep spectrum (CSS) and restarted sources tend to show higher column densities and more disturbed gas kinematics than those in extended radio sources, often exhibiting asymmetric and strongly blueshifted profiles \citep{2015A&A...575A..44G,2017MNRAS.467.2766G}. In our sample of low-power radio sources, we find that extended sources vastly outnumber compact sources. However, previous studies suggest that the intrinsic number of compact radio galaxies exceeds that of extended radio galaxies \citep{2015ApJ...806...59T,2016AN....337...27K}. And the lower fraction of compact sources in our sample may also be influenced by radio power. 

Figure \ref{fig:17} presents the relation between line width and kinematic offset of \hi\ absorption lines and radio power, categorized by radio morphology. From Figure \ref{fig:17}(a), it can be seen that at low radio powers, most absorption lines are relatively narrow. As radio power increases, the number of broad absorption lines also increases, with compact sources dominating in these cases. Combining our sample with that of \cite{2017A&A...604A..43M}, we find that at low radio powers (log($P_{1.4\,\mathrm{GHz}}/\mathrm{W\,Hz}^{-1}$) $<24$), $\sim$26\% of absorbers are offset from the systemic velocity, with redshifted and blueshifted components occurring in roughly equal proportions, most of which are associated with interacting or extended sources. In contrast, at higher radio powers, nearly 50\% of absorbers show velocity offsets, predominantly in compact sources, with blueshifted systems ($\sim$63\%) being dominant. Figure \ref{fig:17}(b) shows that, in low radio power range, most line centroids lie close to the systemic velocity, and a few cases with significant velocity shifts mainly occur in extended and merging sources. With increasing radio power, more sources exhibit velocity offsets in their absorption lines, particularly blueshifted components, which are increasingly dominated by compact sources. These trends suggest that radio power may decrease along the AGN evolutionary path, and the strength of radio jets may play an important role in shaping the interaction between AGNs and ISM.

\subsection{WISE Color-Color Classification}\label{wise}

 \cite{2017A&A...604A..43M} used WISE colors to distinguish between dust-poor sources and MIR-bright galaxies. They found that dust-poor galaxies predominantly exhibit narrow and deep \hi\ absorption lines, typically centered near the systemic velocity, indicating that \hi\ gas is largely settled in rotating disks. In contrast, MIR-bright sources show a higher \hi\ detection rate than dust-poor systems. For sources with $\log(P_{1.4\,\mathrm{GHz}}/\mathrm{W\,Hz^{-1}}) > 24$, these MIR-bright sources often exhibit broad absorption profiles, indicating that \hi\ gas may have disturbed or unstable kinematics.

\begin{figure*}[ht!]
\includegraphics[width=\textwidth]{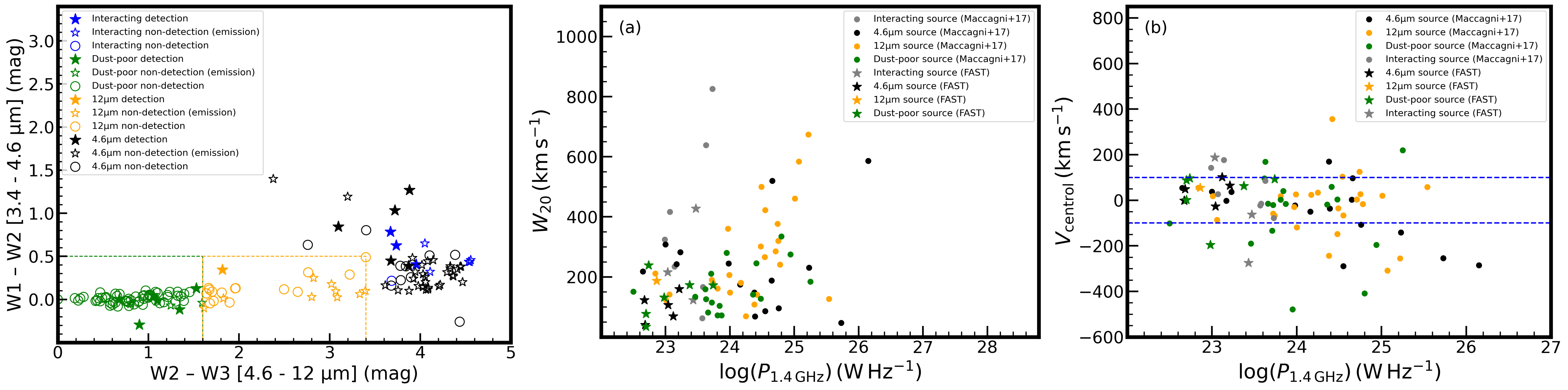}
\caption{The WISE color--color diagram illustrates the MIR properties of the sources in our sample, distinguishing between \hi\ detections (filled symbols) and nondetections (open symbols). Dust-poor sources are shown in green, 12\,$\mu$m bright sources in orange, 4.6\,$\mu$m bright sources in black, and interacting sources in gray. The dashed lines represent the WISE color boundaries used for source classification. Further details on the classification scheme are provided in Table \ref{tab:res}. Sources are classified according to the WISE color. (a) Full width measured at 20\% of the intensity ($W_{20}$) of the \hi\ profiles vs. the radio power of the sources. (b) Line centroid offset with respect to the systemic velocity vs. the radio power of the sources. The fine dashed lines in blue show the interval of $\pm 100\ \text{km s}^{-1}$. } 
\label{fig:wise}
\end{figure*}

Following \cite{2016MNRAS.462.2631M} and \cite{2017A&A...604A..43M}, we classify the sources into several MIR types: dust-poor sources (W1$-$W2 $<$ 0.5 and W2$-$W3 $<$ 1.6); galaxies with MIR emission enhanced by dust continuum, referred to as 12\,$\mu$m bright sources (W1$-$W2 $<$ 0.5 and 1.6 $<$ W2$-$W3 $<$ 3.4); AGN with hot dust in circumnuclear disk and dusty starburst galaxies, referred to as 4.6\,$\mu$m bright sources (the rest regions). In the following sections, we sometimes collectively refer to 12\,$\mu$m bright and 4.6\,$\mu$m bright sources as MIR-bright sources. As shown in Figure \ref{fig:wise} and Table \ref{tab:det}, nearly half of the sources ($\sim44.9\%$) are classified as dust-poor with a detection rate of $9.1^{+4.8}_{-2.4}\%$; $\sim16.3\%$ are 12\,$\mu$m bright sources with a detection rate of $4.2^{+8.4}_{-1.3}\%$; and $\sim31.3\%$ are classified as 4.6\,$\mu$m bright sources with a detection rate of $10.9^{+6.3}_{-3.1}\%$. At low radio powers, considering the uncertainties, the \hi\ absorption detection rate of MIR-bright sources is comparable to that of dust-poor sources. As shown in Figure \ref{fig:wise}(a) and (b), dust-poor sources exhibit narrow and deep \hi\ absorption lines across the full range of radio powers. In contrast, MIR-bright sources at low radio powers also show relatively narrow \hi\ profiles close to the systemic velocity, while those at higher radio powers tend to exhibit broader or more disturbed absorption profiles. Therefore, MIR-bright sources at low radio powers do not show either an enhanced \hi\ absorption detection rate or significantly disturbed \hi\ kinematics. Meanwhile, in the middle panel of Figure \ref{fig:flow}, we find that under the WISE color-color classification, the fractions of inflow and outflow candidates among absorbers show no significant variation across different classes.

\subsection{BPT diagram Classification}\label{bpt}

\begin{figure*}[t!]
\includegraphics[width=\textwidth]{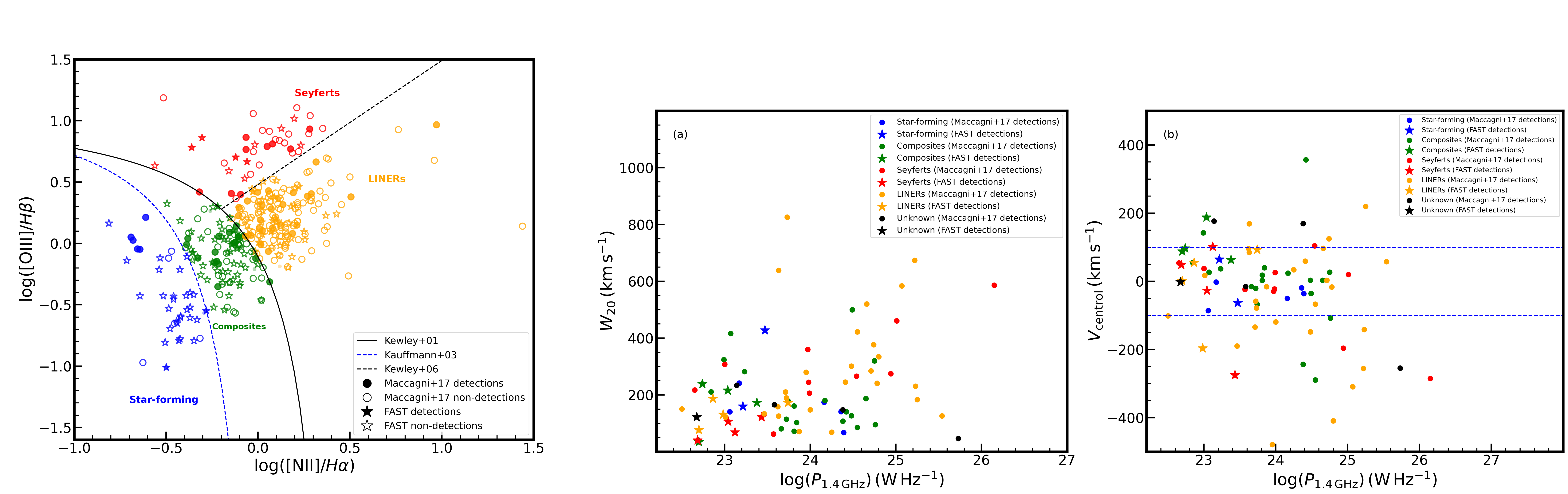}
\caption{The distribution of FAST sample sources and WSRT sample \citep{2017A&A...604A..43M} in the BPT diagram of log([O\,\textsc{iii}]/H$\beta$) versus log([N\,\textsc{ii}]/H$\alpha$) \citep{1981PASP...93....5B}. The solid black curve represents the theoretical demarcation between AGN and composite galaxies defined by \cite{2001ApJ...556..121K}. The dashed blue curve indicates the empirical division between star-forming and composite galaxies proposed by \cite{2003MNRAS.346.1055K}. The dashed black line shows the empirical separation between Seyferts and LINERs defined by \cite{2006MNRAS.372..961K}. Sources are classified according to the BPT diagram. (a) Full width measured at 20\% of the intensity ($W_{20}$) of the \hi\ profiles vs. the radio power of the sources. (b) Line centroid offset with respect to the systemic velocity vs. the radio power of the sources. The fine dashed lines in blue show the interval of $\pm 100\ \text{km s}^{-1}$. 
} 
\label{fig:BPT}
\end{figure*}

\begin{deluxetable*}{lcccc}\setlength\tabcolsep{2.0pt}
\tablenum{3}
\tablecaption{Statistics of the sample in the BPT diagram\label{tab:bpt}}
\tabletypesize{\footnotesize}
\tablewidth{\textwidth}
\tablehead{
\colhead{Sample} &\colhead{BPT Classification} & \colhead{Number of Sources} & \colhead{Detections} & \colhead{Detection rate (\%)}
}
\decimalcolnumbers
\startdata
{} & Seyfert & 29 & 10 & $34.5^{+9.6}_{-7.6}\%$ \\
{} & LINERs & 125 & 24 & $19.2^{+4.0}_{-3.0}\%$ \\
WSRT sample & Composites & 45 & 18 & $40.0^{+7.6}_{-6.7}\%$ \\
{} & Star-forming & 12 & 5 & $41.7^{+14.4}_{-12.0}\%$ \\
{} & Unkown & 37 & 9 & \nodata \\
\hline
{} & Seyfert & 11 & 4 & $36.4^{+15.5}_{-11.3}\%$ \\
{} & LINERs & 42 & 4 & $9.5^{+6.5}_{-2.8}\%$ \\
FAST sample & Composites & 52 & 4 & $7.7^{+5.4}_{-2.3}\%$ \\
{} & Star-forming & 31 & 2 & $6.5^{+7.4}_{-2.2}\%$ \\
{} & Unkown & 11 & 1 & \nodata \\
\hline
{} & Seyfert & 40 & 14 & $35.0^{+8.1}_{-6.7}\%$ \\
{} & LINERs & 167 & 28 & $16.8^{+3.3}_{-2.5}\%$ \\
WSRT+FAST sample & Composites & 97 & 22 & $22.7^{+4.8}_{-3.7}\%$ \\
{} & Star-forming & 43 & 7 & $16.3^{+7.1}_{-4.2}\%$ \\
{} & Unkown & 48 & 10 & \nodata
\enddata
 \tablecomments{Column (1): the name of samples. Column (2): the classification method of the sources. Column (3): the number of observed source. Column (4): the number of \hi\ absorption detections. Column (5): the detection rates in each subsample.}

\end{deluxetable*}

We classify both our FAST sample and the WSRT sample from \cite{2017A&A...604A..43M} using the BPT diagram. The detailed classification results are summarized in Figure \ref{fig:BPT} and Table \ref{tab:bpt}. In Table \ref{tab:bpt}, we compare and list the detection rates of \hi\ absorption lines in each subsample characterized by the BPT diagram. The left panel of Figure \ref{fig:BPT} presents the classification of radio sources in the combined sample based on AGN activity using the BPT diagram, while Figure \ref{fig:BPT}(a) and (b) compare the spectral profiles and kinematics of \hi\ absorption, as classified in the BPT diagram, with radio power. We find that a small fraction of the radio sources are classified as star forming, where the radio continuum is primarily powered by supernova-driven starbursts. Moreover, the fraction of star-forming galaxies in the FAST sample ($\sim21\%$) is slightly higher than that in the WSRT sample ($\sim5\%$). Considering the \hi\ in star-forming sources may have different origins and properties, we tested the impacts of star-forming dominated sources by excluding them in both samples, and found no significant changes to the statistical results of the \hi\ absorbers. In the right panel of Figure \ref{fig:flow}, we show the distribution of absorbers (total of FAST sample and WSRT sample) across different classes under BPT diagram classification, together with the number and fraction of infalling/outflow candidates. 

As shown in Figure \ref{fig:BPT}, most of the absorbers in our low-power radio FAST sample exhibit narrow line widths and velocities close to the systemic galaxies. In contrast, a substantial fraction of the sources in the higher-power WSRT sample of \cite{2017A&A...604A..43M} show broad absorption features with large velocity offsets. This absorber is more commonly found in Seyferts and LINERs.

To investigate how the absorber properties in different BPT classes vary with radio power, we combine the two samples for a direct comparison. Among the detected absorbers, 28 sources ($\sim35.4^{+5.7}_{-4.9}\%$ of absorbers) exhibit redshifted or blueshifted features ($|\Delta v| > 100~\mathrm{km\,s^{-1}}$), and are therefore considered candidates of infalling/outflowing gas. Combining Figure \ref{fig:BPT}(a) and (b), and the right panel of Figure \ref{fig:flow}, we find that absorbers in star forming galaxies are predominantly associated with low radio powers ($\log(P_{\rm 1.4\,GHz}/{\rm W\,Hz^{-1}}) < 24$), characterized by narrow widths close to the systemic velocity, likely regular in rotating disks. No redshifted or blueshifted absorbers are identified in star forming galaxies, while only six such cases are found in Composites, corresponding to a low fraction, and none are present at high radio powers. And in different AGN types, the fractions of inflow/outflow candidates do not show a significant difference, with a slightly higher outflow rate in LINERs and Seyferts. With increasing radio power, absorbers in Seyferts and LINERs display broader line widths, suggestive of disturbed gas, accompanied by a growing number and fraction of redshifted/blueshifted systems. Moreover, Figure \ref{fig:BPT}(b) clearly shows that blueshifted absorbers become dominant at higher radio powers, while at lower powers ($\log(P_{\rm 1.4\,GHz}/{\rm W\,Hz^{-1}}) < 24$), blueshifted features are exclusively associated with Seyferts and LINERs. It indicates that Seyferts and LINERs are more efficient in driving gas and interacting with ISM, and that their activity strengthens with increasing radio power.

\section{Summary}\label{sec:sum}

In the study, we conduct a search for associated \hi\ absorption in low-power radio sources ($\log(P_{\rm 1.4\,GHz}/{\rm W\,Hz^{-1}}) < 24$) at low redshift ($z < 0.1$), using the Five-hundred-meter Aperture Spherical Telescope (FAST). This work extends the survey of \cite{2023ApJ...952..144Y}, with a total of 159 radio sources observed. Seven sources are unusable due to severe RFI,  and the final sample comprises 147 radio sources. We detected 12 new absorbers, and combining our results with those of \cite{2023ApJ...952..144Y}, 
we identify 15 \hi\ absorbers (see Table \ref{tab:res}, Table \ref{tab:a1}, and \ref{sec:B}). We examine the detection rate, origin, and kinematics of the absorbing \hi\ gas, classifying the sample by radio morphology (compact vs.\ extended), WISE colors (dust-poor, 12\,$\mu$m-bright, and 4.6\,$\mu$m-bright), and BPT diagrams. Our aim is to investigate the origin of absorbing gas 
and the impact of radio power on the interaction between AGNs and interstellar medium (ISM). The main results are summarized as follows:

\begin{enumerate}
    \item $10.2^{+3.1}_{-2.0}\%$ of radio galaxies in our sample exhibit associated \hi\ absorption. Compared to sources with higher-power radio at similar redshifts, low-power radio AGNs show a relatively lower detection rate of \hi\ absorption. The low detection rate may be attributed to dilution from \hi\ emission and differences in radio morphology. It may also suggest that these sources have less active SMBH at lower radio power ($\log(P_{1.4\,\mathrm{GHz}}/\mathrm{W\,Hz^{-1}}) < 24$).
    \item In low-power radio sources ($\log(P_{1.4\,\mathrm{GHz}}/\mathrm{W\,Hz}^{-1})<24$), most \hi\ absorption lines are narrow and centered near systemic velocity, consistent with rotating disks. Broad lines tracing disturbed \hi\ components induced by radio activity are only found in merging galaxies, implying that regular disks dominate in low-power radio systems. However, a few absorbers show significant velocity offsets, indicative of jet-driven outflows, infalling \hi\ clouds, or HVCs. The fraction of outflow candidates increases with radio power, while inflow candidates remain constant, suggesting the radio emission of AGN may have effects on driving \hi\ outflows.
    \item Compact sources exhibit a higher \hi\ detection rate compared to extended sources. In the low-power radio sources ($\log(P_{1.4\,\mathrm{GHz}}/\mathrm{W\,Hz}^{-1}) < 24$), extended sources dominate our sample. This implies a gas-rich and star-forming-dominated population in low-power sources. 
    \item Dust-poor sources exhibit narrow and deep \hi\ absorption lines across the entire range of radio powers. MIR-bright sources at low radio powers do not show either a high \hi\ absorption detection rate or evidence of disturbed \hi\ kinematics.
    \item Among different AGN types, the fractions of inflow and outflow candidates show slightly higher outflow rates in LINERs and Seyferts. With increasing radio power, absorbers in Seyferts and LINERs show broader profiles and a higher incidence of velocity-shifted systems, with blueshifted absorbers becoming dominant. At low-power radio sources, blueshifted absorption is detected only in Seyferts and LINERs, indicating the connection between atomic outflows and the ionization state of AGN, and their activity may progressively enhance with increasing radio power.

\end{enumerate}

\begin{acknowledgments}
We thank the referee for careful reading and suggestions that improved the paper. This work is supported by the National SKA Program of China 2025SKA0150103, and the National Natural Science Foundation of China under grant Nos. 11890692, 12133008, 12221003, and 12373011. Q.Y. was supported by the European Research Council (ERC) under grant agreement No. 101040751. J.W. acknowledges the National Key R{\&}D Program of China (grant No. 2023YFA1607904) and the NSFC grant 12333002. B.Z. is also supported by  Guizhou Provincial Science and Technology Projects (Nos.QKHFQ[2023]003, QKHFQ[2024]001, and QKHPTRC-ZDSYS[2023]003). We acknowledge the science research grants from the China Manned Space Project, grant Nos. CMS-CSST-2021-A04, CMS-CSST-2021-A05, CMS-CSST-2021-A06, CMS-CSST-2021-B02, and CMS-CSST-2025-A10.  

This work made use of the data from FAST. FAST, a Chinese national mega-science facility, operated by the National Astronomical Observatories, Chinese Academy of Sciences.
\end{acknowledgments}

\bibliography{ref}{}
\bibliographystyle{aasjournal}



\appendix

\onecolumngrid
\section{Summary of Absorbers}\label{sec:B}

\begin{figure*}[ht!]
\centering
\includegraphics[width=\textwidth]{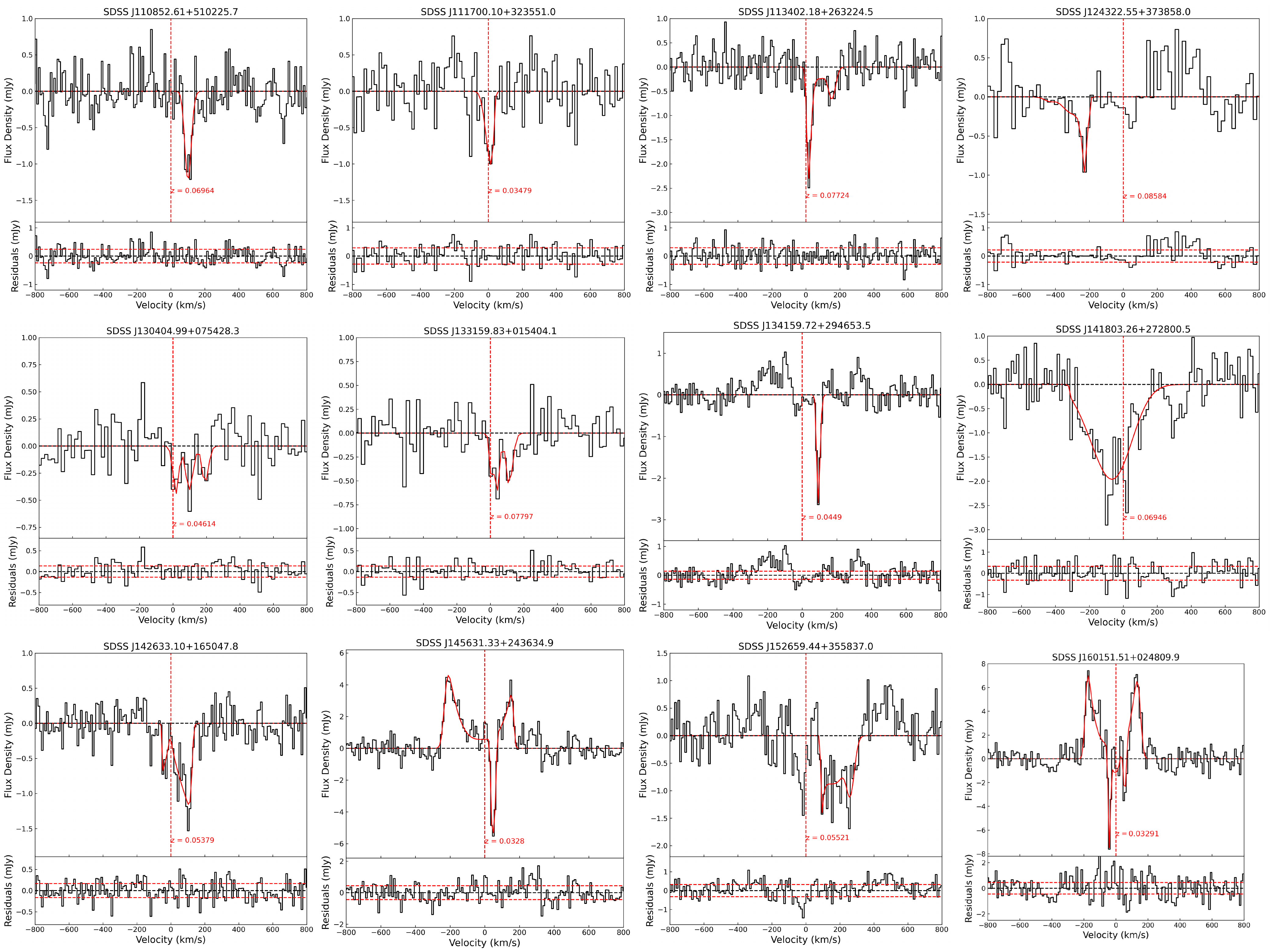} 
 \caption{\hi\ absorption detections. The black solid line shows the absorption spectrum, and the red solid line represents the Gaussian model fit. The red dotted-dashed vertical line marks the redshift provided by SDSS spectroscopy \citep{2020ApJS..249....3A}. The residual of the fit is shown at the bottom, with the $\pm1\sigma$ noise level indicated by red dashed lines.}

\label{fig:absorbers}
\end{figure*}

The \hi\ absorption spectra in FAST sample with fitted profiles are shown in Figure \ref{fig:absorbers}.

\onecolumngrid
\section{Summary Table of Nondetections}

We summarized the ancillary information of the \hi\ absorption nondetections in Table \ref{tab:a1}.

\startlongtable
\begin{deluxetable*}{lccccccccccc}

\tablenum{B1}
\tablecaption{\hi\ Absorption Nondetections\label{tab:a1}}
\tabletypesize{\scriptsize}
\tablewidth{\textwidth}
\setlength{\tabcolsep}{3pt}
\tablehead{
\colhead{Source Name} & \colhead{R.A.} & \colhead{Decl.} & \colhead{$z$} & \colhead{$S_{\text{1.4\ GHz, NVSS}}$} & \colhead{$S_{\text{1.4\ GHz, FIRST}}$} & \colhead{log($P_{\text{1.4\ GHz}}$)} & \colhead{rms} & \colhead{$\tau_{\text{peak}}$} & \colhead{Radio Type} & \colhead{WISE Type} & \colhead{Other Name}\\
\colhead{} & \colhead{(deg)} & \colhead{(deg)}  & \colhead{} & \colhead{(mJy)} & \colhead{(mJy $\text{beam}^{-1}$)} & \colhead{(W $\text{Hz}^{-1}$)} & \colhead{(mJy)} & \colhead{} & \colhead{} & \colhead{} & \colhead{}
}
\decimalcolnumbers
\startdata
SDSS J104808.12+181119.2 & 162.03 & 18.19 & 0.02002 & 22.2 & 21.3 & 22.31 & 0.25 & $<$0.036 & E & dp & \nodata\\
SDSS J105233.59+452417.3 & 163.14 & 45.40 & 0.08962 & 25.5 & \nodata & 23.73 & 0.37 & \nodata & E & dp & \nodata\\
SDSS J105359.84+493934.9 & 163.50 & 49.66 & 0.02208 & 10.4 & 2.2 & 22.07 & 0.23 & $<$0.368 & E & 12um & \nodata\\
SDSS J110436.95+450730.7 & 166.15 & 45.13 & 0.02164 & 17.4 & 2.3 & 22.27 & 0.40 & $<$0.741 & E & 4.6um & \nodata\\
SDSS J110508.11+444447.1 & 166.28 & 44.75 & 0.02155 & 15.6 & 5.2 & 22.22 & 0.46 & $<$0.310 & E & 4.6um & \nodata\\
SDSS J110801.50+211905.0 & 167.01 & 21.32 & 0.07886 & 14.1 & 3.9 & 23.35 & 0.26 & $<$0.224 & E & dp & \nodata\\
SDSS J110801.87+310946.4 & 167.01 & 31.16 & 0.07473 & 10.0 & 11.2 & 23.15 & 0.24 & $<$0.066 & C & dp & \nodata\\
SDSS J110937.43+265515.6 & 167.41 & 26.92 & 0.07030 & 23.3 & 16.5 & 23.46 & 0.27 & $<$0.051 & E & 4.6um & \nodata\\
SDSS J111005.84+152441.2 & 167.52 & 15.41 & 0.08116 & 10.8 & 11.0 & 23.26 & 0.17 & $<$0.048 & E & 4.6um & \nodata\\
SDSS J111246.75+132401.2 & 168.19 & 13.40 & 0.06845 & 10.0 & 1.5 & 23.07 & 0.18 & $<$0.466 & E & dp & \nodata\\
SDSS J111248.53+253550.5 & 168.20 & 25.60 & 0.05049 & 10.3 & 4.8 & 22.80 & 0.15 & $<$0.099 & E & 4.6um & \nodata\\
SDSS J111257.49+301028.3 & 168.24 & 30.17 & 0.02942 & 10.5 & 11.8 & 22.33 & 0.16 & $<$0.041 & E & 4.6um & \nodata\\
SDSS J111750.61+263732.8 & 169.46 & 26.63 & 0.02702 & 10.1 & 7.4 & 22.23 & 0.19 & $<$0.078 & C & 12um & \nodata\\
SDSS J112055.83+173854.0 & 170.23 & 17.65 & 0.08451 & 21.2 & 2.6 & 23.59 & 0.36 & $<$0.538 & E & dp & \nodata\\
SDSS J112230.06+241645.3 & 170.63 & 24.28 & 0.02963 & 24.2 & 3.7 & 22.69 & 0.28 & $<$0.253 & E & 12um & \nodata\\
SDSS J112239.02+374554.4 & 170.66 & 37.77 & 0.00664 & 22.9 & \nodata & 21.35 & 0.30 & \nodata & E & 4.6um & \nodata\\
SDSS J112253.54+342027.3 & 170.72 & 34.34 & 0.03551 & 14.8 & 10.2 & 22.64 & 0.27 & $<$0.081 & E & 12um & \nodata\\
SDSS J112333.92+022348.5 & 170.89 & 2.40 & 0.07547 & 27.3 & 22.8 & 23.60 & 0.40 & $<$0.054 & E & dp & \nodata\\
SDSS J112617.75+465826.8 & 171.57 & 46.97 & 0.02489 & 15.3 & 4.3 & 22.34 & 0.32 & $<$0.254 & E & 4.6um & \nodata\\
SDSS J112946.69+315916.7 & 172.44 & 31.99 & 0.07343 & 19.1 & 2.4 & 23.42 & 0.27 & $<$0.418 & E & dp & \nodata\\
SDSS J113221.83+004816.7 & 173.09 & 0.80 & 0.01996 & 11.1 & 1.2 & 22.00 & 0.20 & $<$0.647 & E & 4.6um & \nodata\\
SDSS J113355.19+024254.4 & 173.48 & 2.72 & 0.07483 & 10.8 & 7.3 & 23.19 & 0.15 & $<$0.063 & E & dp & \nodata\\
SDSS J113446.55+485721.9 & 173.69 & 48.96 & 0.03157 & 13.4 & 36.6 & 22.49 & 0.24 & $<$0.020 & E & dp & IC~711\\
SDSS J113910.98+392002.0 & 174.80 & 39.33 & 0.02376 & 10.6 & 5.0 & 22.14 & 0.16 & $<$0.101 & E & 4.6um & \nodata\\
SDSS J113922.34+321327.7 & 174.84 & 32.22 & 0.07362 & 10.9 & 1.8 & 23.18 & 0.21 & $<$0.421 & E & dp & \nodata\\
SDSS J114740.08+334720.4 & 176.92 & 33.79 & 0.03090 & 16.4 & 15.0 & 22.56 & 0.27 & $<$0.056 & C & 12um & \nodata\\
SDSS J114804.60+372638.1 & 177.02 & 37.44 & 0.04176 & 28.9 & 28.3 & 23.08 & 0.33 & $<$0.036 & E & dp & \nodata\\
SDSS J114907.37+110033.6 & 177.28 & 11.01 & 0.08608 & 11.0 & 2.0 & 23.32 & 0.14 & $<$0.242 & E & dp & \nodata\\
SDSS J115019.17+105009.6 & 177.58 & 10.84 & 0.08516 & 14.7 & 11.8 & 23.44 & 0.19 & $<$0.049 & E & 12um & \nodata\\
SDSS J115105.27+482835.1 & 177.77 & 48.48 & 0.06729 & 21.4 & 12.2 & 23.38 & 0.34 & $<$0.087 & E & 12um & \nodata\\
SDSS J115147.62+484059.3 & 177.95 & 48.68 & 0.00324 & 12.2 & 2.1 & 20.45 & 0.27 & $<$0.496 & E & 4.6um & NGC~3928\\
SDSS J115224.06+322413.9 & 178.10 & 32.40 & 0.01023 & 14.3 & 13.7 & 21.53 & 0.19 & $<$0.042 & C & 4.6um & \nodata\\
SDSS J115339.96+432739.3 & 178.42 & 43.46 & 0.01969 & 19.4 & \nodata & 22.24 & 0.62 & \nodata & I & 4.6um & \nodata\\
SDSS J115538.34+430245.1 & 178.91 & 43.05 & 0.02366 & 16.9 & 4.8 & 22.34 & 0.25 & $<$0.173 & E & 4.6um & \nodata\\
SDSS J115827.53+153101.9 & 179.61 & 15.52 & 0.06917 & 10.4 & 9.5 & 23.10 & 0.13 & $<$0.043 & E & dp & \nodata\\
SDSS J120416.77+240849.0 & 181.07 & 24.15 & 0.05101 & 21.4 & 13.3 & 23.13 & 0.35 & $<$0.082 & E & dp & \nodata\\
SDSS J120515.56+431008.4 & 181.31 & 43.17 & 0.05285 & 21.5 & 19.4 & 23.17 & 0.39 & $<$0.062 & E & dp & \nodata\\
SDSS J120615.88+245655.7 & 181.57 & 24.95 & 0.07533 & 14.4 & 10.4 & 23.32 & 0.16 & $<$0.048 & E & 12um & \nodata\\
SDSS J120802.07+094557.0 & 182.01 & 9.77 & 0.06942 & 14.0 & 3.0 & 23.23 & 0.37 & $<$0.462 & E & dp & \nodata\\
SDSS J121748.47+463452.7 & 184.45 & 46.58 & 0.05744 & 15.5 & 7.2 & 23.10 & 0.35 & $<$0.158 & E & 4.6um & \nodata\\
SDSS J122024.05+253338.0 & 185.10 & 25.56 & 0.02289 & 28.3 & 18.3 & 22.53 & 0.57 & $<$0.098 & E & dp & \nodata\\
SDSS J123104.55+285114.8 & 187.77 & 28.85 & 0.06108 & 28.2 & 21.0 & 23.42 & 0.38 & $<$0.055 & I & dp & \nodata\\
SDSS J124651.21+001749.3 & 191.71 & 0.30 & 0.08917 & 14.5 & 5.4 & 23.48 & 0.34 & $<$0.211 & E & dp & \nodata\\
SDSS J124958.51+060943.6 & 192.49 & 6.16 & 0.06847 & 11.9 & \nodata & 23.15 & 0.16 & \nodata & E & 12um & \nodata\\
SDSS J125607.37+132156.4 & 194.03 & 13.37 & 0.06848 & 10.0 & 8.1 & 23.07 & 0.25 & $<$0.098 & E & 4.6um & \nodata\\
SDSS J130256.62+354005.8 & 195.74 & 35.67 & 0.03708 & 16.2 & 13.5 & 22.72 & 0.32 & $<$0.074 & I & 4.6um & \nodata\\
SDSS J130551.95+414318.9 & 196.47 & 41.72 & 0.02384 & 16.8 & 2.2 & 22.34 & 0.25 & $<$0.418 & E & 4.6um & \nodata\\
SDSS J130600.43+434608.4 & 196.50 & 43.77 & 0.03906 & 10.8 & 9.5 & 22.59 & 0.18 & $<$0.059 & E & 4.6um & \nodata\\
SDSS J131146.38+402353.5 & 197.94 & 40.40 & 0.07094 & 10.6 & 3.8 & 23.13 & 0.12 & $<$0.097 & E & dp & \nodata\\
SDSS J131325.85+433214.7 & 198.36 & 43.54 & 0.05752 & 24.3 & 20.5 & 23.30 & 0.51 & $<$0.077 & E & 4.6um & \nodata\\
SDSS J131517.26+442425.5 & 198.82 & 44.41 & 0.03553 & 14.3 & 11.1 & 22.63 & 0.26 & $<$0.072 & I & 4.6um & \nodata\\
SDSS J131840.21+313234.5 & 199.67 & 31.54 & 0.03593 & 11.2 & 5.6 & 22.53 & 0.18 & $<$0.101 & E & 4.6um & \nodata\\
SDSS J132012.74+171223.1 & 200.05 & 17.21 & 0.08482 & 27.9 & 28.3 & 23.72 & 0.57 & $<$0.062 & C & dp & \nodata\\
SDSS J132102.61+332022.1 & 200.26 & 33.34 & 0.03643 & 10.0 & 5.1 & 22.50 & 0.18 & $<$0.109 & E & dp & \nodata\\
SDSS J132348.44+431804.2 & 200.95 & 43.30 & 0.02735 & 10.0 & 9.6 & 22.24 & 0.25 & $<$0.081 & E & 4.6um & \nodata\\
SDSS J132543.85+171552.7 & 201.43 & 17.26 & 0.07877 & 11.2 & 3.6 & 23.25 & 0.14 & $<$0.126 & E & 4.6um & \nodata\\
SDSS J132803.96+341957.7 & 202.02 & 34.33 & 0.03409 & 10.4 & 3.9 & 22.45 & 0.15 & $<$0.125 & E & dp & \nodata\\
SDSS J132951.93+402758.3 & 202.47 & 40.47 & 0.08113 & 22.1 & 7.0 & 23.57 & 0.35 & $<$0.162 & E & dp & \nodata\\
SDSS J133248.70+415218.5 & 203.20 & 41.87 & 0.02728 & 13.0 & \nodata & 22.35 & 0.38 & \nodata & E & 4.6um & \nodata\\
SDSS J133935.95+430310.0 & 204.90 & 43.05 & 0.01180 & 10.9 & 1.8 & 21.53 & 0.31 & $<$0.748 & E & 4.6um & \nodata\\
SDSS J135732.92+104714.0 & 209.39 & 10.79 & 0.06427 & 11.1 & \nodata & 23.06 & 0.13 & \nodata & E & 12um & \nodata\\
SDSS J135814.74+182046.2 & 209.56 & 18.35 & 0.06274 & 21.1 & 3.5 & 23.31 & 0.24 & $<$0.228 & E & dp & \nodata\\
SDSS J135907.04+393736.1 & 209.78 & 39.63 & 0.07927 & 16.0 & 10.5 & 23.41 & 0.22 & $<$0.066 & E & dp & \nodata\\
SDSS J140246.51+051057.1 & 210.69 & 5.18 & 0.08126 & 17.1 & 16.9 & 23.46 & 0.22 & $<$0.040 & C & dp & \nodata\\
SDSS J140458.34+383803.2 & 211.24 & 38.63 & 0.06505 & 11.0 & 10.7 & 23.06 & 0.17 & $<$0.050 & E & dp & \nodata\\
SDSS J140621.57+504330.2 & 211.59 & 50.73 & 0.00638 & 29.0 & 1.8 & 21.42 & 0.47 & $<$1.541 & E & 4.6um & NGC~5480\\
SDSS J140904.22+415640.3 & 212.27 & 41.94 & 0.06469 & 13.5 & 6.2 & 23.15 & 0.21 & $<$0.105 & E & 4.6um & \nodata\\
SDSS J141617.09+511646.0 & 214.07 & 51.28 & 0.07404 & 14.2 & 5.5 & 23.30 & 0.28 & $<$0.168 & E & dp & \nodata\\
SDSS J141918.81+394035.8 & 214.83 & 39.68 & 0.01957 & 18.5 & 21.1 & 22.21 & 0.42 & $<$0.061 & C & 4.6um & \nodata\\
SDSS J142041.03+015930.8 & 215.17 & 1.99 & 0.07831 & 11.7 & 3.7 & 23.26 & 0.20 & $<$0.179 & E & 12um & \nodata\\
SDSS J142336.57+482610.2 & 215.90 & 48.44 & 0.07400 & 11.2 & 11.3 & 23.19 & 0.19 & $<$0.051 & E & dp & \nodata\\
SDSS J142436.26+024442.5 & 216.15 & 2.75 & 0.05413 & 10.2 & 11.9 & 22.86 & 0.15 & $<$0.039 & E & dp & \nodata\\
SDSS J142447.40+023951.9 & 216.20 & 2.66 & 0.05351 & 11.2 & 10.3 & 22.89 & 0.14 & $<$0.042 & C & dp & \nodata\\
SDSS J142453.68+170216.4 & 216.22 & 17.04 & 0.05342 & 24.7 & 1.5 & 23.24 & 0.37 & $<$1.273 & E & dp & \nodata\\
SDSS J142728.39+460847.6 & 216.87 & 46.15 & 0.00774 & 27.9 & 1.2 & 21.57 & 0.47 & \nodata & E & 4.6um & \nodata\\
SDSS J142819.23+291844.2 & 217.08 & 29.31 & 0.08700 & 11.7 & 12.0 & 23.36 & 0.31 & $<$0.081 & E & dp & \nodata\\
SDSS J143923.52+404800.8 & 219.85 & 40.80 & 0.01049 & 18.4 & 18.5 & 21.66 & 0.37 & $<$0.062 & C & 4.6um & \nodata\\
SDSS J143953.88+100754.6 & 219.97 & 10.13 & 0.06244 & 17.8 & 11.1 & 23.24 & 0.36 & $<$0.103 & E & dp & \nodata\\
SDSS J144003.44+370727.4 & 220.01 & 37.12 & 0.09765 & 14.9 & 4.4 & 23.58 & 0.30 & $<$0.228 & E & 12um & \nodata\\
SDSS J144057.04+463647.2 & 220.24 & 46.61 & 0.04521 & 12.3 & 9.8 & 22.78 & 0.24 & $<$0.076 & E & 4.6um & \nodata\\
SDSS J144524.57+384647.0 & 221.35 & 38.78 & 0.03259 & 13.4 & 11.6 & 22.52 & 0.21 & $<$0.057 & I & 4.6um & \nodata\\
SDSS J144609.68+191721.0 & 221.54 & 19.29 & 0.04382 & 11.8 & 12.0 & 22.73 & 0.28 & $<$0.073 & E & 12um & \nodata\\
SDSS J145132.52+114043.7 & 222.89 & 11.68 & 0.06305 & 11.3 & 6.0 & 23.05 & 0.17 & $<$0.091 & E & 4.6um & \nodata\\
SDSS J145927.63+163842.3 & 224.87 & 16.65 & 0.03342 & 16.8 & 4.9 & 22.64 & 0.26 & $<$0.170 & E & dp & \nodata\\
SDSS J150453.80+231523.7 & 226.22 & 23.26 & 0.04584 & 26.8 & 18.5 & 23.13 & 0.29 & $<$0.047 & E & dp & \nodata\\
SDSS J150628.54+462039.0 & 226.62 & 46.34 & 0.03640 & 12.9 & 2.9 & 22.61 & 0.23 & $<$0.267 & E & 12um & \nodata\\
SDSS J150928.17+073324.5 & 227.37 & 7.56 & 0.07719 & 27.5 & 32.4 & 23.62 & 0.39 & $<$0.037 & C & dp & \nodata\\
SDSS J151131.90+180618.4 & 227.88 & 18.11 & 0.08125 & 10.2 & 3.3 & 23.24 & 0.26 & $<$0.274 & E & dp & \nodata\\
SDSS J151315.56+403107.5 & 228.31 & 40.52 & 0.03177 & 10.0 & 5.5 & 22.37 & 0.18 & $<$0.104 & E & 4.6um & \nodata\\
SDSS J151338.41+210738.5 & 228.41 & 21.13 & 0.07886 & 26.7 & 5.5 & 23.63 & 0.44 & $<$0.271 & E & dp & \nodata\\
SDSS J151903.50+042001.1 & 229.76 & 4.33 & 0.04681 & 11.2 & 3.6 & 22.77 & 0.15 & $<$0.131 & E & dp & \nodata\\
SDSS J152139.97+254845.9 & 230.42 & 25.81 & 0.07164 & 10.4 & 7.7 & 23.13 & 0.13 & $<$0.053 & E & dp & \nodata\\
SDSS J152527.48+050029.9 & 231.36 & 5.01 & 0.03575 & 16.4 & 10.9 & 22.69 & 0.20 & $<$0.058 & E & 4.6um & \nodata\\
SDSS J152949.83+033039.1 & 232.46 & 3.51 & 0.03687 & 16.7 & 8.2 & 22.73 & 0.24 & $<$0.094 & E & 12um & \nodata\\
SDSS J153016.15+270551.0 & 232.57 & 27.10 & 0.03255 & 13.3 & 13.1 & 22.52 & 0.19 & $<$0.044 & E & dp & \nodata\\
SDSS J153033.65+014752.5 & 232.64 & 1.80 & 0.08234 & 11.2 & 8.9 & 23.29 & 0.27 & $<$0.096 & C & 12um & \nodata\\
SDSS J153713.26+431753.9 & 234.31 & 43.30 & 0.01914 & 23.9 & 1.4 & 22.30 & 0.58 & \nodata & E & 4.6um & \nodata\\
SDSS J153827.56+062333.1 & 234.61 & 6.39 & 0.06691 & 11.4 & 3.8 & 23.11 & 0.22 & $<$0.188 & E & dp & \nodata\\
SDSS J154206.27+483514.3 & 235.53 & 48.59 & 0.06799 & 12.2 & 3.9 & 23.15 & 0.26 & $<$0.219 & E & dp & \nodata\\
SDSS J155636.39+415250.5 & 239.15 & 41.88 & 0.03471 & 14.8 & 7.3 & 22.62 & 0.44 & $<$0.199 & E & 4.6um & \nodata\\
SDSS J161209.28+000333.1 & 243.04 & 0.06 & 0.05801 & 22.5 & 16.7 & 23.27 & 0.40 & $<$0.075 & C & dp & \nodata\\
SDSS J161238.84+293836.8 & 243.16 & 29.64 & 0.03209 & 27.2 & 22.0 & 22.82 & 0.39 & $<$0.054 & C & dp & \nodata\\
SDSS J162415.17+201100.7 & 246.06 & 20.18 & 0.03973 & 10.3 & 5.6 & 22.59 & 0.20 & $<$0.115 & E & 4.6um & \nodata\\
SDSS J162952.88+242638.4 & 247.47 & 24.44 & 0.03787 & 22.2 & 18.9 & 22.88 & 0.37 & $<$0.061 & I & 4.6um & MRK~883\\
SDSS J163741.47+304147.0 & 249.42 & 30.70 & 0.05531 & 22.3 & 23.1 & 23.22 & 0.34 & $<$0.045 & C & dp & \nodata\\
SDSS J164145.23+393835.2 & 250.44 & 39.64 & 0.01631 & 10.1 & \nodata & 21.79 & 0.12 & \nodata & E & 4.6um & \nodata\\
SDSS J164540.67+214919.0 & 251.42 & 21.82 & 0.03201 & 11.6 & 6.4 & 22.44 & 0.21 & $<$0.102 & E & 4.6um & \nodata\\
SDSS J164909.57+361325.8 & 252.29 & 36.22 & 0.03097 & 26.9 & 25.8 & 22.78 & 0.32 & $<$0.038 & E & 12um & \nodata\\
SDSS J213046.02+103727.4 & 322.69 & 10.62 & 0.06177 & 18.5 & 12.0 & 23.24 & 0.40 & $<$0.105 & E & dp & \nodata\\
SDSS J031202.49-000442.5\tablenotemark{a} & 48.01 & $-$0.08 & 0.03762 & 14.7 & 7.5 & 22.68 & 0.19 & $<$0.079 & E & dp & \nodata\\
SDSS J032232.78-000004.2\tablenotemark{a} & 50.64 & $-$0.00 & 0.02175 & 17.2 & 6.7 & 22.27 & 0.18 & $<$0.084 & E & 4.6um & \nodata\\
SDSS J035212.57-062056.2\tablenotemark{a} & 58.05 & $-$6.35 & 0.03295 & 12.5 & 7.4 & 22.50 & 0.22 & $<$0.094 & E & \nodata & \nodata\\
SDSS J040643.39-050653.0\tablenotemark{a} & 61.68 & $-$5.11 & 0.06600 & 14.2 & \nodata & 23.18 & 0.21 & \nodata & E &  \nodata & \nodata\\
SDSS J073555.67+421212.1\tablenotemark{a} & 113.98 & 42.20 & 0.08822 & 18.5 & 17.3 & 23.56 & 0.20 & $<$0.035 & C & 12um & \nodata\\
SDSS J074632.46+302926.7\tablenotemark{a} & 116.64 & 30.49 & 0.05704 & 16.3 & \nodata & 23.10 & 0.20 & \nodata & E & dp & \nodata\\
SDSS J083950.75+230836.1\tablenotemark{a} & 129.96 & 23.14 & 0.02520 & 12.3 & 10.7 & 22.25 & 0.15 & $<$0.043 & C & 4.6um & \nodata\\
SDSS J085022.80+552243.4\tablenotemark{a} & 132.60 & 55.38 & 0.03076 & 20.0 & 10.2 & 22.64 & 0.30 & $<$0.092 & E & dp & \nodata\\
SDSS J090302.14+402602.3\tablenotemark{a} & 135.76 & 40.43 & 0.02877 & 15.5 & 13.8 & 22.47 & 0.18 & $<$0.040 & C & 12um & \nodata\\
SDSS J092002.16+010217.8\tablenotemark{a} & 140.01 & 1.04 & 0.01703 & 10.7 & 11.4 & 21.84 & 0.17 & $<$0.046 & E & 12um & \nodata\\
SDSS J102219.38+363458.9\tablenotemark{a} & 155.58 & 36.58 & 0.02592 & 13.4 & 7.7 & 22.31 & 0.24 & $<$0.099 & E & 12um & \nodata\\
SDSS J110637.35+460219.6\tablenotemark{a} & 166.66 & 46.04 & 0.02501 & 16.3 & 14.6 & 22.37 & 0.23 & $<$0.049 & E & 4.6um & \nodata\\
SDSS J112039.95+504938.2\tablenotemark{a} & 170.17 & 50.83 & 0.02762 & 24.8 & 15.0 & 22.64 & 0.40 & $<$0.084 & E & dp & \nodata\\
SDSS J112458.71+470836.6\tablenotemark{a} & 171.24 & 47.14 & 0.05372 & 10.5 & 8.7 & 22.86 & 0.23 & $<$0.083 & C & 12um & \nodata\\
SDSS J114047.35+463225.7\tablenotemark{a} & 175.20 & 46.54 & 0.05359 & 14.9 & 17.6 & 23.01 & 0.18 & $<$0.031 & E & dp & \nodata\\
SDSS J115330.58+524121.9\tablenotemark{a} & 178.38 & 52.69 & 0.07164 & 23.0 & \nodata & 23.46 & 0.45 & \nodata & E & 12um & \nodata\\
SDSS J134056.58+262912.1\tablenotemark{a} & 205.24 & 26.49 & 0.07502 & 26.0 & \nodata & 23.55 & 0.25 & \nodata & C & dp & \nodata\\
SDSS J142351.53+401531.8\tablenotemark{a} & 215.96 & 40.26 & 0.08220 & 12.1 & 8.1 & 23.31 & 0.15 & $<$0.058 & E & dp & \nodata\\
SDSS J144545.10+513450.9\tablenotemark{a} & 221.44 & 51.58 & 0.02963 & 11.5 & 11.2 & 22.37 & 0.19 & $<$0.052 & C & 4.6um & \nodata\\
SDSS J153929.67+443854.4\tablenotemark{a} & 234.87 & 44.65 & 0.07299 & 20.6 & 18.9 & 23.43 & 0.32 & $<$0.052 & C & dp & \nodata\\
SDSS J154451.23+433050.6\tablenotemark{a} & 236.21 & 43.51 & 0.03664 & 11.4 & 12.1 & 22.55 & 0.14 & $<$0.035 & E & dp & \nodata\\
SDSS J162423.55+250748.4\tablenotemark{a} & 246.10 & 25.13 & 0.09603 & 12.6 & 10.3 & 23.47 & 0.21 & $<$0.063 & E & dp & \nodata\\
SDSS J212811.65+001759.5\tablenotemark{a} & 322.05 & 0.30 & 0.05250 & 15.2 & 6.5 & 23.00 & 0.18 & $<$0.086 & E & dp & \nodata
\enddata
\tablecomments{Columns (1): the source name. Columns (2): R.A. in degrees. Columns (3): Decl. in degrees. Columns (4): redshift. Columns (5): radio continuum flux at 1.4 GHz from NVSS. Columns (6): radio continuum peak flux at 1.4 GHz from FIRST \citep{2015ApJ...801...26H}}. Columns (7): radio power at 1.4 GHz. Columns (8): 1$\sigma$ noise measured at the velocity resolution of 10 $\text{km s}^{-1}$. Columns (9): upper limit of the peak optical depth estimated with 3$\sigma$ rms. Columns (10): radio morphology classification as compact (C) and extended (E). Columns (11): WISE classification type. Columns (12): Other names of the radio source.
\tablenotetext{a}{The detailed data and measurements of this source are adopted from \cite{2023ApJ...952..144Y}.}
\end{deluxetable*}

\section{Notes on some radio sources in the FAST sample}\label{C}

\noindent \textbf{SDSS J113446.55+485721.9}: Also known as IC 711. The radio galaxy IC 711 has been observed and studied because it exhibits a narrow-angle-tail morphology \citep{2020MNRAS.493.3811S}.

\noindent \textbf{SDSS J115147.62+484059.3}: Also known as NGC 3928. NGC 3928 is a blue compact early-type starburst galaxy that has been extensively observed in earlier studies \citep[e.g.,][]{1987ApJ...313...89T,1990ApJ...350L..29G,1994AJ....107...90L}.

\noindent \textbf{SDSS J115339.96+432739.3}: Its DESI image reveals prominent gaseous and stellar tidal tails, indicating that it has recently undergone or is currently experiencing an interaction or merger event. It has also been classified as a merging system in previous study \citep{2024A&A...686A.151C}. Therefore, we consider it to be an interacting source. The \hi\ spectrum of this source  also shows a detection of \hi\ emission.

\noindent \textbf{SDSS J123104.55+285114.8}: The DESI image shows that it is a double-nuclei galaxy hosting a dual AGN system, which has been studied \citep{2023MNRAS.524.4482B}. This indicates that the galaxy is in a merging stage; therefore, we classify this galaxy as an interacting source.

\noindent \textbf{SDSS J124322.55+373858.0}: The DESI image reveals that the galaxy exhibits a prominent gaseous and stellar tail, indicating that it has recently undergone or is currently undergoing a merging event. It has also been studied as a merging system in previous work \citep{2019MNRAS.487.5490P}; therefore, we classify it as an interacting source.

\noindent \textbf{SDSS J130256.62+354005.8}: The DESI image clearly shows that the galaxy possesses a large gaseous and stellar tail. In addition, its color index satisfies $u-r < 2$. These features suggest that the galaxy has recently undergone, or is currently undergoing, a merging event. Therefore, we classify this galaxy as an interacting source. The \hi\ spectrum of this source  also shows a detection of \hi\ emission.

\noindent \textbf{SDSS J131517.26+442425.5}: The DESI image shows that the galaxy is merging with a spiral companion. In previous study \citep{2022ApJS..261...34H}, it has also been investigated as a merging system. Therefore, we classify it as an interacting source. The \hi\ spectrum of this source  also shows a detection of \hi\ emission.

\noindent \textbf{SDSS J140621.57+504330.2:} It also known as NGC~5480 and is classified as an SA(s)c spiral galaxy with a redshift of $z = 0.00638$. As a galaxy that has hosted multiple supernova explosions, NGC~5480 has been the subject of extensive observational studies \citep{1988IAUC.4590....1P,2022ApJ...927...61K}.

\noindent \textbf{SDSS J141803.26+272800.5}: Its DESI image clearly shows large gaseous and stellar tidal tails, and it has been classified as a merging galaxy in previous study \citep{2017A&A...602A.100T}. Therefore, we classify it as an interacting source.

\noindent \textbf{SDSS J144524.57+384647.0}: The DESI image clearly shows that the galaxy has prominent gaseous and stellar tidal tails. In addition, its color index satisfies $u-r < 2$. These features indicate that the galaxy has recently undergone, or is currently undergoing, a merging event; therefore, we classify it as an interacting source. The \hi\ spectrum of this source  also shows a detection of \hi\ emission.

\noindent \textbf{SDSS J152659.44+355837.0}: The DESI image reveals extensive gas and tidal tails, indicating that the galaxy is in a merging stage. It has also been classified as a merging system in previous study \citep{2024A&A...686A.151C}. Therefore, we classify this source as an interacting galaxy.

\noindent \textbf{SDSS J162952.88+242638.4:} It is also known as Mrk~883, which is a nearby Seyfert galaxy at a redshift of $z = 0.03787$ and hosts a double-peaked narrow emission line AGN. In the DESI image, Mrk~883 exhibits a clearly disturbed morphology, featuring a gaseous ring and tidal tails, indicative of a recent merger event. This source has been recently observed and reported in detail by \cite{2024Univ...10...21B}.

All \hi\ spectra are presented and can be accessed at \url{https://github.com/Tsuki348/hi-spectra.git}.

\end{CJK*}
\end{document}